\documentclass[preprint,endfloats,showpacs,preprintnumbers,amsmath,amssymb]{revtex4}

\usepackage{psfig}

\begin{document}


\title{Theory of domain patterns in systems with long-range
interactions of Coulomb type} 

\author{C. B. Muratov}
\email{muratov@njit.edu}
\affiliation{Department of Mathematical Sciences, New Jersey Institute
of Technology, Newark, NJ 07102}

\date{\today}

\begin{abstract}

We develop a theory of the domain patterns in systems with competing
short-range attractive interactions and long range repulsive Coulomb
interactions. We take an energetic approach, in which patterns are
considered as critical points of a mean-field free energy
functional. Close to the microphase separation transition, this
functional takes on a universal form, allowing to treat a number of
diverse physical situations within a unified framework. We use
asymptotic analysis to study domain patterns with sharp interfaces. We
derived an interfacial representation of the pattern's free energy
which remains valid in the fluctuating system, with a suitable
renormalization of the Coulomb interaction's coupling constant. We
also derived integrodifferential equations describing the stationary
domain patterns of arbitrary shapes and their thermodynamic stability,
coming from the first and second variation of the interfacial free
energy. We showed that the length scale of a stable domain pattern
must obey a certain scaling law with the strength of the Coulomb
interaction. We analyzed existence and stability of localized (spots,
stripes, annuli) and periodic (lamellar, hexagonal) patterns in two
dimensions. We showed that these patterns are metastable in certain
ranges of the parameters and that they can undergo morphological
instabilities leading to the formation of more complex patterns. We
discuss nucleation of the domain patterns by thermal fluctuations and
pattern formation scenarios for various thermal quenches. We argue
that self-induced disorder is an intrinsic property of the domain
patterns in the systems under consideration.

\end{abstract}


\maketitle

\newcommand{\ep}{\epsilon}
\newcommand{\al}{\alpha}
\newcommand{\phb}{\bar\phi}
\newcommand{\rs}{{\cal R}}
\newcommand{\ls}{{\cal L}}
\newcommand{\rb}{\bar\rho}

\bibliographystyle{apsrev}

\section{Introduction} 

Pattern formation is a beautiful example of cooperative behavior in
complex systems. It is most pronounced in open dissipative systems
maintained away from thermal equilibrium by external fluxes of energy
or matter
\cite{cross93,mikhailov,murray,kapral,niedernostheide,ko:book}. At the
same time, there exists a great number of systems interacting with the
outside world only through the contact with a heat bath which are also
capable of pattern formation and self-organization. Typically, these
systems are characterized by the presence of coexisting phases, or a
phase transition which is the driving force for the cooperative
behavior. Examples of such classical systems include ferroelectric and
ferromagnetic films, ferrofluids, Langmuir monolayers, various polymer
systems, etc. (see, for example,
\cite{odell,setter,grosberg,bates90,bates99,seul95}). Among such
quantum systems are type-I superconductors in the intermediate state,
high-temperature superconductors, degenerate ferromagnetic
semiconductors, etc. (see, for example
\cite{huebener,care75,nagaev95,kovalev86}).

In systems not far from thermal equilibrium patterns may form as a
result of the competition of interactions operating on different
length scales \cite{seul95}. Typically, a short-range attractive
interaction in the system would favor macroscopic phase
separation. The latter, however, is counteracted by a long-range
repulsive interaction. This is often accompanied by a microphase
separation transition, that leads to spontaneous formation of patterns
in the ideally homogeneous systems upon variations of the control
parameters.

An important class of systems with competing interactions are systems
in which the long-range interaction is Coulombic. The fundamental
nature of the Coulomb interaction makes this class of systems
extremely diverse. These systems include a variety of polymer systems,
such as block copolymers
\cite{bates90,bates99,leibler80,ohta86,liu89}, weakly charged
polyelectrolyte solutions \cite{erukhimovich93,borue88,nyrkova94},
cross-linked polymer mixtures \cite{degennes79}; amphiphile solutions
\cite{stillinger83}; phase-separating ceramic compounds \cite{chen93};
systems undergoing reaction-controlled spinodal decomposition
\cite{glotzer95}; photostimulated phase transitions
\cite{ko:book,kovalev86,mamin94}, etc. Some aspects of systems with
competing interactions are shared by systems far from thermal
equilibrium, such as heated electron-hole and gas plasma,
semiconductor devices \cite{ko:book,niedernostheide}, crystal surfaces
undergoing laser-induced melting \cite{yeung94}, autocatalytic
chemical reactions and surface catalytic reactions
\cite{kapral,hildebrand99}. Furthermore, a number of quantum systems,
such as degenerate magnetic semiconductors and high-temperature
superconductors which exhibit electronic phase separation can be
considered as systems with competing Coulomb interactions
\cite{nagaev95,emery93}. In addition, the general problem of Wigner
crystallization \cite{care75,lundqvist}, as well as the thermodynamic
and glassy properties of spin systems frustrated by Coulomb
interaction
\cite{emery93,low94,viot98,grousson01,lorenzana01,schmalian00}, can be
considered from this point of view.

Here we develop a theory of patterns with sharp interfaces (domains)
in systems with short-range attractive interactions and long-range
repulsive Coulomb interactions. Our starting point is a mean-field
free energy functional, which has a nonlocal term associated with the
Coulomb interaction. Specifically, we are interested in the case of a
weak Coulomb interaction, when domain patterns with sharp interfaces
are realized. We view patterns as critical points of the free energy
functional. Our main tool in the analysis is singular perturbation
theory based on the strong separation of length scales in the systems
under consideration. We use the results of our analysis of the domain
patterns to study nucleation and formation of complex patterns. We
also discuss the effect of thermal fluctuations and thermodynamic
properties of these systems.

Our paper is organized as follows. In Sec. II, we introduce the
general free energy functional and its reduction near a ``local''
critical point, derive the interfacial representation of the free
energy and develop a renormalization scheme to account for the effect
of thermal fluctuations. In Sec. III we derive the asymptotic
equations for the stationary patterns and their stability. In Sec. IV
we perform a detailed analysis of localized and periodic patterns in
two dimensions. In Sec. V we discuss nucleation and growth of complex
patterns as a result of instabilities of simple patterns, and in
Sec. VI we draw conclusions. This paper is partially based on the
author's Ph. D. Thesis \cite{m:phd}.

\section{Systems with competing interactions of Coulomb type}

\subsection{Free energy functional}

We start by considering the following general mean-field free energy
functional
\begin{eqnarray} \label{FF}
F = \int d^d x \left( {|\nabla \phi|^2 \over 2} + f(\phi) \right. &&
 \nonumber \\ && \hspace{-4cm} \left. + {\al \over 2} \int d^d x'
 g[\phi(x)] G(x - x') g[\phi(x')] \right). \label{F:gen}
\end{eqnarray}
Here $\phi(x)$ is a scalar order parameter, $f(\phi)$ is a double-well
potential, $G(x - x')$ is a positive-definite long-range kernel, $\al$
is a (positive) coupling constant, $g(\phi)$ is a monotone function
that is equal to zero at some $\phi = \phb$, and $d$ is the
dimensionality of space. Here and henceforth we use dimensionless
units.

The functional in Eq.~(\ref{FF}) may be applicable to a variety of
systems. Generally, $\phi$ may stand for magnetization, density of the
charged polymer in a polyelectrolyte solution, volume fraction of a
block copolymer in a diblock copolymer melt, density of electrons or
holes in a charge density wave, structural state of a catalytic
surface, concentration of a chemical species, etc.
\cite{nagaev95,borue88,leibler80,ohta86,mcmillian75,mamin94,%
hildebrand99,emery93}. The kernel $G(x - x')$ we are interested in is
the {\em Coulomb} potential, i.e., it satisfies
\begin{eqnarray}
 \label{G}
 - \nabla^2 G(x - x') = \delta^{(d)} (x - x'),
\end{eqnarray}
where $\delta^{(d)}(x)$ is the $d$-dimensional Dirac
delta-function. The physical nature of the Coulomb interaction may
also significantly vary from system to system: it may arise as a
result of the actual electrostatic repulsion due to charges associated
with the order parameter, it may have an entropic origin, as in block
copolymers, or it can come from the diffusion of chemically reacting
species (see, for example,
\cite{borue88,mcmillian75,leibler80,chen93,ohta86,mamin94,ko:book,%
emery93,glotzer95,yeung94,hildebrand99}). Note that in quantum systems
Eq.~(\ref{FF}) arises within the framework of density functional
theory \cite{hohenberg64,smith69,lundqvist}.

The long-ranged nature of $G(x - x')$ from Eq.~(\ref{G}) is expressed
in the fact that its Fourier transform has a singularity at wave
vector $\mathbf k = \mathbf 0$. At the same time, this Fourier
transform is bounded at large $k$-vectors. Let us emphasize that $G(x
- x')$ represents a {\em repulsive} long-range interaction, since it
is positive for all $x$. Therefore, the Coulomb long-range interaction
represented by $G(x - x')$ is {\em competing} with the short-range
interactions represented by the first two terms in
Eq.~(\ref{F:gen}). It is also clear that since the Fourier transform
of $G(x - x')$ is positive for all wave vectors, the functional in
Eq.~(\ref{F:gen}) is bounded from below on any finite domain.

If we formally put $\al = 0$ in Eq.~(\ref{F:gen}), we will recover the
standard free energy functional that is used in the studies of phase
separation (see, for example, \cite{bray94}). On the other hand,
no matter how small the value of $\alpha$ is, because of the
singularity of the Fourier-transform of $G(x - x')$ at $\mathbf k =
\mathbf 0$ the effect of the long-range interaction will remain
significant on sufficiently large length scales. Indeed, if the system
has a finite size $L$, from the dimensional considerations the
contribution of the Coulomb interaction into the free energy will
scale as $\al L^{d + 2}$. If the value of $\al$ is decreased while $L$
remains fixed, the contribution of the Coulomb interaction goes
away. This means that when $\al \ll 1$, the system behaves locally as
if it did not have the long-range interaction. On the other hand, for
an infinite system this interaction is always relevant since its
contribution scales as $L^{d + 2} > L^d$.  Therefore, for
\begin{eqnarray}
 \label{al} \al \ll 1
\end{eqnarray}
the long-range interaction will be a {\em singular perturbation},
globally affecting the behavior of the system. It is in this case that
domain patterns form in systems with the free energy of the form of
Eq.~(\ref{F:gen}). Since we are interested in the domain patterns
here, Eq.~(\ref{al}) will be assumed from now on. Note that this
condition is satisfied in many systems with long-range interactions of
Coulomb type
\cite{ohta86,leibler80,borue88,stillinger83,degennes79,nagaev95,emery93}.

The singularity of $G(x - x')$ on the large length scales implies that
the Fourier component of $g(\phi)$ at $\mathbf k = \mathbf 0$ must
vanish in order for the last integral in Eq.~(\ref{F:gen}) to remain
finite. This corresponds to the overall electroneutrality for systems
in which the order parameter is associated with the electric
charge. The only possible {\em homogeneous} phase of the system is,
therefore, $\phi = \phb$. Thus, due to the long-range interaction the
global phase separation in the system becomes impossible. On the other
hand, as we will see below, the system described by the free energy
functional from Eq.~(\ref{F:gen}) may be in a {\em patterned}
state. By patterned states (more precisely, by stationary patterns),
we will mean the inhomogeneous distributions of the order parameter
which are critical points of the functional $F$.

\subsection{The microphase separation transition} \label{s:mf}

Let us assume that in the absence of the long-range interaction the
system would possess a critical point at temperature $T = T_c$.  Then,
in the vicinity of $T_c$ the function $f(\phi)$ can be expanded as
\begin{eqnarray} \label{lg:gen}
 \label{f} f \simeq { a \tau \phi^2 \over 2} + {b \phi^4 \over 4},
\end{eqnarray}
where $\tau = (T - T_c) / T_c$ is the reduced temperature, and $a$ and
$b$ are positive constants \cite{landau5}. In the following, we will
talk about $T_c$ as the ``local'' critical temperature. Near $T_c$,
the value of $|\phi| \sim \phi_0 = (a |\tau| /b)^{1/2} \ll 1$
\cite{landau5}. If also $|\phb| \ll 1$, we can expand the function
$g(\phi)$ in a Taylor series and retain only the first term, so
$g(\phi) \simeq {\rm const} \times (\phi - \phb)$. Then, rescaling the
order parameter and length with the values of $\phi_0$ and the
short-range correlation length $\xi = |a \tau|^{-1/2}$ \cite{landau5},
we can write the free energy from Eq. (\ref{F:gen}) below $T_c$ in the
following {\em universal} form:
\begin{eqnarray}
 \label{F} F = \int d^d x \left( {|\nabla \phi|^2 \over 2} - {\phi^2
 \over 2} + {\phi^4 \over 4} \right. && \nonumber \\ && \left.
 \hspace{-5cm} + {\ep^2 \over 2} \int d^d x' {(\phi(x) - \phb) G(x -
 x') (\phi(x') - \phb)} \right),
\end{eqnarray}
where we absorbed a constant factor into the definition of $F$.
Here the parameter $\ep$, which plays the role of the effective
coupling constant of the long-range interaction, is given by
\begin{eqnarray}
 \label{ep} \ep = \al^{1/2} \left| {g'(0) \over a \tau} \right| \sim
\al^{1/2} | \tau |^{-1}.
\end{eqnarray}
Notice that in Eq.~(\ref{F}) the value of $\phb$ has been rescaled as
well, so it now depends on temperature:
\begin{eqnarray} \label{scale:gen}
\phb \propto |\tau|^{-1/2}.
\end{eqnarray}
Also, as was discussed above, for Eq.~(\ref{F}) the singularity of
$G(x - x')$ at small wave vectors implies that the total amount of the
order parameter must be conserved (the ``electroneutrality''
condition):
\begin{eqnarray} \label{conserv}
{1 \over V} \int \phi \, d^d x = \phb,
\end{eqnarray}
where $V$ is the system's volume.

Let us consider small fluctuations of the order parameter $\delta \phi
= \phi - \phb$ away from the homogeneous phase for $T < T_c$. From the
second variation of $F$ from Eq. (\ref{F}), the Fourier transform of
the pair correlation function of such fluctuations is
\begin{eqnarray}
 \label{dphi} \langle | \delta \phi_\mathbf{k}|^2 \rangle \propto {V
\over |\mathbf k|^2 + 3 \phb^2 - 1 + \ep^2 |\mathbf k|^{-2} }.
\end{eqnarray}
This correlation function has a maximum at non-zero $k$-vectors with
$|\mathbf k| = k_c$, where
\begin{eqnarray}
 \label{kc} k_c = \ep^{1/2}.
\end{eqnarray}
The fluctuations at $k_c$ diverge when
$\phb = \pm |\phb_c|$, where
\begin{eqnarray}
 \label{phic} |\phb_c| = \frac{1}{\sqrt{3}} \left( 1 - {\ep \over
\ep_c} \right)^{1/2}, ~~~  \ep_c = {1 \over 2}.
\end{eqnarray}

The divergence of the fluctuations at $|\mathbf k| = k_c$ signifies an
instability of the homogeneous phase and leads to the {\em microphase
separation} \cite{seul95}. Note that the instability can only be
realized if $\ep$ is small enough; in terms of temperature, it occurs
at $T$ slightly below $T_c$ when Eq.~(\ref{al}) holds, see
Eq.~(\ref{ep}).

As the temperature is decreased, the value of $\ep$ gets smaller. Note
that for small $\al$ one can still be close to $T_c$ and yet have $\ep
\ll 1$. In this situation the long-range interaction can be a singular
perturbation (in the sense discussed earlier) even in the vicinity of
the transition. As was already mentioned, this is a necessary
condition for the existence of the domain patterns, so below we will
concentrate on the case $\ep \ll 1$.  For $\ep \ll 1$ the instability
of the homogeneous phase occurs close to the classical spinodal of the
Ginzburg-Landau free energy: $|\phb_c| \simeq 1 / \sqrt{3}$, see
Eq.~(\ref{phic}). In this case, according to Eq. (\ref{kc}), the
instability occurs at $k_c \ll 1$.

There are two regions in the $k$-space in which the fluctuations of
the order parameter around the homogeneous phase, with $|\phb| >
|\phb_c|$, behave differently when $\ep \ll 1$. According to
Eq. (\ref{dphi}), for $|\mathbf k| \sim 1$ one could neglect the
long-range contribution, so the fluctuations $\langle |\delta
\phi_\mathbf{k}|^2 \rangle \propto V /( |\mathbf k|^2 + m^2)$, where
$m^2 = 3 \phb^2 - 1$, are those of the (mean-field) critical phenomena
\cite{landau5}, with the length scale independent of $\ep$:
\begin{eqnarray}
l \sim 1.
\end{eqnarray}
On the other hand, for $|\mathbf k| \ll 1$ one can neglect the
$|\mathbf k|^2$ term, so the fluctuations behave like $\langle |
\delta \phi |_\mathbf{k}^2 \rangle \propto V / ( m^2 + \ep^2 |\mathbf
k|^{-2}) = (V / m^2) \left[ 1 - \ep^2 / ( \ep^2 + m^2 |\mathbf k|^2 )
\right]$. The first term in this expression represents local order
parameter fluctuations, while the second is the familiar
Debye-H\"{u}ckel correlation function \cite{landau5}. The length scale
associated with the latter is the {\em screening} length
\begin{eqnarray}
L \sim \ep^{-1}.
\end{eqnarray}

For $\ep \ll 1$ the (generally, metastable) equilibrium state of the
system should be a {\em domain pattern} made up of domains of large
size $\sim R$ separated by narrow domain walls of width $\sim
l$. Clearly, the long-range interaction cannot significantly affect
the local profiles of the order parameter; however, it can affect the
{\em locations} of the domain walls. The size of the domains will be
determined by the competition between the surface energy of the domain
walls $\sim R^{d-1}$ per droplet and the energy of the long-range
interaction $\sim \ep^2 R^{d + 2}$, so the characteristic size of the
equilibrium domain pattern will be of order (in the context of block
copolymers, see also \cite{ohta86})
\begin{eqnarray}
 \label{R:gen} R \sim \ep^{-2/3}.
\end{eqnarray}
Note that this result for the global minimizers of the sharp interface
limit of Eq.~(\ref{F}) was recently proved by Choksi
\cite{choksi01}. Choksi also obtained rigorous upper and lower bounds
on the energy of global minimizers of Eq.~(\ref{F}) in the situation
when the screening effects are negligible.

According to Eq.~(\ref{kc}), the wavelength of the fluctuations with
respect to which the instability of the homogeneous phase is realized
is
\begin{eqnarray}
\lambda = 2 \pi/k_c \sim \ep^{-1/2}.
\end{eqnarray}
Comparing all these length scales, one can see that for $\ep \ll 1$
the following hierarchy holds:
\begin{eqnarray}
 \label{hier:gen}
 l \ll \lambda \ll R \ll L.
\end{eqnarray}
This is a crucial property of systems with weak long-range Coulomb
interaction.

\subsection{Interfacial representation of the free energy}
\label{s:int} 

The solutions of the Euler-Lagrange equation obtained from
Eq.~(\ref{F}) may be analyzed by singular perturbations theory in the
asymptotic limit $\ep \rightarrow 0$. We will perform this analysis in
Sec. \ref{s:stat}.  Now, however, we will use a different method which
gives the free energy of the domain pattern in terms of the locations
of the domain interfaces \cite{m:phd}. This method was used by
Goldstein, Muraki, and Petrich for a reaction-diffusion system with a
weak activator-inhibitor coupling \cite{petrich94,goldstein96}. Here
we develop a procedure that allows to calculate the free energy of a
domain pattern which takes into account the screening effects.

Because of the strong separation of length scales we can introduce the
following {\em ansatz} for the distribution of the order parameter:
\begin{eqnarray}
 \label{shsm} \phi(x) = \phi_\mathrm{sh}(x) + \phi_\mathrm{sm}(x),
\end{eqnarray}
where $\phi_\mathrm{sh}$ represents the sharp distributions, whose
characteristic length of variation is of the order of the domain wall
width (which in our units is of order one), and $\phi_\mathrm{sm}$
represents the smooth distributions, whose characteristic length of
variation is comparable to the domain size $R$. The distribution
$\phi_\mathrm{sh}$ is chosen in such a way that it is equal to $+1$
inside the positive domains and $-1$ outside, whereas at the
interfaces it is close to the one-dimensional domain wall of the
Ginzburg-Landau theory \cite{landau8}: 
\begin{eqnarray} \label{sh}
\phi_\mathrm{sh} = \tanh {\rho \over \sqrt{2}},
\end{eqnarray}
where $\rho$ is the distance from a given point to the interface,
which is positive (negative) in the positive (negative) domains,
respectively. Thus, the location of the interface is built into the
definition of $\phi_\mathrm{sh}$. The contribution from
$\phi_\mathrm{sh}$ to the free energy, coming from the integration in
Eq.~(\ref{F}) in the vicinity (of order 1) of the interfaces, gives
the surface energy
\begin{eqnarray}
 \label{interf} F_\mathrm{surf} = \sigma_0 \oint d S, ~~~\sigma_0 = {2
 \sqrt{2} \over 3}.
\end{eqnarray}
Here the surface integral gives the total surface area of the domain
interfaces and $\sigma_0$ is the surface tension coefficient of the
domain wall in the Ginzburg-Landau theory \cite{landau8}.

To find the smooth distributions $\phi_\mathrm{sm}$ away from the
interfaces, we minimize the free energy in these regions. Taking into
account that $\phi_\mathrm{sm}$ varies slowly on the length scale of
order 1, we can neglect the $\nabla^2 \phi$ term arising in the
Euler-Lagrange equation and obtain
\begin{eqnarray}
 \label{euler:sm} \mu + \phi - \phi^3 - \ep^2 \int d^d x' G(x - x')
 (\phi(x') - \phb) = 0,
\end{eqnarray}
where $\mu$ is the chemical potential (a constant) coming from the
constraint given by Eq.~(\ref{conserv}).  On the scale of the domains,
$\phi_\mathrm{sh} = \pm 1$ away from the interfaces. We will assume
that inside the domains $|\phi_\mathrm{sm}| \ll 1$, which is justified
for $R \ll \ep^{-1}$ (see below). This allows us to linearize
Eq. (\ref{euler:sm}) around $\phi_\mathrm{sh}$ away from the
interfaces. Using Eq.  (\ref{shsm}) with $\phi_\mathrm{sh}^2 = 1$,
Eq.~(\ref{euler:sm}) is written as
\begin{eqnarray}
 \label{lc} \phi_\mathrm{sm} = - \kappa^2 \psi, ~~~\kappa^2 =
 \frac{1}{2},
\end{eqnarray}
where we introduced an effective field
\begin{eqnarray}
 \label{psi} \psi = -\mu + \ep^2 \int d^d x' G(x - x')
 (\phi_\mathrm{sh}(x') + \phi_\mathrm{sm}(x') - \phb). \nonumber \\
\end{eqnarray}
Note that the constant $\kappa^2$ is basically the coefficient of
linear response for the local theory. 

Applying $\nabla^2$ to Eq.~(\ref{psi}) and using Eqs. (\ref{G}) and
(\ref{lc}), we obtain
\begin{eqnarray}
 \label{psiscreened} - \nabla^2 \psi + \ep^2 \kappa^2 \psi = \ep^2
(\phi_\mathrm{sh} - \phb).
\end{eqnarray}
Note that our definition of $\psi$, together with Eq.~(\ref{lc}),
implies that Eq.~(\ref{conserv}) is automatically satisfied to the
leading order in $\ep$. This can be seen by integrating
Eq.~(\ref{psiscreened}) over the volume of the system and taking into
account that for no-flux or periodic boundary conditions the surface
integral in the obtained expression vanishes. Also note that for the
same reason $\mu$ drops out from this equation.

The solution of Eq.~(\ref{psiscreened}) is
\begin{eqnarray}
 \label{sm} \psi = \ep^2 \int d^dx' G_\ep  (x - x')
 (\phi_\mathrm{sh}(x') - \phb),
\end{eqnarray}
where $G_\ep $ is the screened Coulomb interaction that satisfies
\begin{eqnarray}
 \label{Gsc}
 - \nabla^2 G_\ep  + \ep^2 \kappa^2 G_\ep  = \delta^{(d)} (x - x').
\end{eqnarray}
These are explicitly given as follows
\begin{eqnarray}
 \label{Gsc123}
 G_\ep  (x - x') =  \left\{ 
\begin{array}{ll}
{1 \over 2 \ep \kappa} \exp(-\ep \kappa |x - x'|) & \mathrm{in~~} d =
1, \\ \\ {1 \over 2 \pi} K_0 ( \ep \kappa |x - x'|) & \mathrm{in~~} d
= 2, \\ \\ {\exp( - \ep \kappa |x - x'|) \over 4 \pi |x - x'|} &
\mathrm{in~~} d = 3,
\end{array}
\right.
\end{eqnarray}
where $K_0(x)$ is the modified Bessel function. Thus, the fluctuations
of the order parameter in the bulk indeed screen the interaction on
the length scale $L \sim \ep^{-1}$. This means that the finite size
effects will become unimportant if the system size is much greater
than $L$. Notice that the value of $\psi$ is estimated as $\psi \sim
\ep^2 R^2 \ll 1$ for $R \ll \ep^{-1}$, justifying the linearization
used in the derivation. Also, according to Eqs.~(\ref{lc}) and
(\ref{shsm}), in this situation the deviation of $\phi$ from $\pm 1$
is small away from the interfaces.

Let us now calculate the contribution from the long-range interaction
to the free energy. Once again, neglecting the $|\nabla \phi|^2$ term,
expanding the nonlinearity in Eq.~(\ref{F}) around $\phi_\mathrm{sh}$
up to the second order in $\phi_\mathrm{sm}$, and taking into account
that $\phi_\mathrm{sh}^2 = 1$ away from the interfaces, to the leading
order in $\ep$ we can write the contribution of the long-range
interaction (up to an overall constant) as follows:
\begin{eqnarray}
 \label{longcontr} F_\mathrm{long-range} = && \nonumber \\ &&
 \hspace{-2.5cm} \int d^d x \left( {1 \over 2 \kappa^2}
 \phi_\mathrm{sm}^2 + \frac{1}{2} (\phi_\mathrm{sh} -\phb) (\psi +
 \mu) + \frac{1}{2} \phi_\mathrm{sm} (\psi + \mu) \right) \nonumber \\
 && \hspace{-2cm} = {\ep^2 \over 2} \int d^d x d^d x'
 (\phi_\mathrm{sh}(x) - \phb) G_\ep (x - x') (\phi_\mathrm{sh}(x') -
 \phb), \nonumber \\
\end{eqnarray}
where we used Eqs.~(\ref{conserv}), (\ref{lc}), (\ref{psi}), and
(\ref{sm}). One can see from this equation that the screening
represented by $\phi_\mathrm{sm}$ enters the free energy only via
$G_\ep(x - x')$. The integral in Eq.~(\ref{longcontr}) can be
transformed to an integral over the domain interfaces by using
Eq. (\ref{Gsc}) and the fact that $\phi_\mathrm{sh} = \pm 1$ in the
positive (negative) domains \cite{goldstein96}. After calculating the
respective integrals and collecting all the terms in the free energy
(see Appendix \ref{a:free}), we obtain
\begin{eqnarray}
 F = \sigma_0 \oint dS - {2 (1 + \phb) \over \kappa^2} \int_{\Omega_+}
 d^d x \nonumber \\ && \hspace{-4cm} + 2 \ep^2 \int_{\Omega_+}
 \int_{\Omega_+} d^d x d^d x' G_\ep (x - x') \label{F:int} \\ =
 \sigma_0 \oint dS - {2 \phb \over \kappa^2 d} \oint dS (\vec{x} \cdot
 \hat{n}) && \nonumber \\ && \hspace{-4cm} - \frac{2}{\kappa^2} \oint
 dS \oint dS' (\hat{n} \cdot \hat{n}') G_\ep (x - x'), \label{F:inter}
\end{eqnarray}
where $\Omega_+$ denotes the positive domains, $\hat{n}$ is the
outward normal to the interface of $\Omega_+$, and the surface
integrals are over the interface. The first integral in
Eq.~(\ref{F:inter}) is the overall surface area of the interfaces, the
second gives the total volume of the positive domains, and the third
is the non-local contribution of the screened long-range interaction
(note the distinction with \cite{goldstein96}). Thus,
Eq.~(\ref{F:inter}) gives the free energy of the domain pattern in
terms of the locations of the interfaces only. Note that the
unscreened version of Eq.~(\ref{F:int}) was recently derived
rigorously by Ren and Wei in the context of $\Gamma$-convergence
\cite{ren00}.

\subsection{Renormalization} \label{s:renorm}

The treatment above is based on the mean-field free energy functional
from Eq.~(\ref{F}) and therefore neglects the effects of thermal
fluctuations. In a fluctuating theory, this functional will become an
effective Hamiltonian, in general requiring an appropriate
field-theoretic treatment. Nevertheless, we propose that the effect of
the critical phenomena fluctuations can be taken into account by an
appropriate renormalization of the main parameters of the free energy
in the interfacial regime \cite{m1:prl97}.

Indeed, if one looks at the singularly perturbed [Eq.~(\ref{al})]
fluctuating system near $T_c$, on small length scales one will see
critical phenomena fluctuations of a second-order phase transition
{\em without} the long-range interaction. This will happen as long as
the characteristic screening length $L$ of the long-range interaction
is much greater than the correlation length $\xi$ of the critical
phenomena fluctuations. The critical exponents associated with the
local critical phenomena fluctuations must be those of the
$d$-dimensional Ising model \cite{landau5}. So, the local average of
the order parameter will be close to a constant $\phi = \pm \phi_0
|\tau|^\beta$, where $\tau = (T - T_c)/T_c$ is the reduced temperature
and $\beta$ is the respective critical exponent. Also, the surface
tension coefficient of an interface in which the order parameter
changes sign is $\sigma = \sigma_0 |\tau|^{\nu (d - 1)}$, where $\nu$
is another critical exponent, and its width is roughly the correlation
length $\xi = \xi_0 |\tau|^{-\nu}$ \cite{landau5}.

Observe that the long-range coupling involves integration over regions
of size $\sim R$ which for the domain patterns must be much greater
than the correlation length. Therefore, it is the {\em average} value
of the order parameter that gives the main contribution to the
long-range interaction energy for $R \gg \xi$.  This energy has to be
compared with the surface energy, so in equilibrium we obtain
\begin{eqnarray}
 \label{comp:ren} \sigma_0 |\tau|^{\nu(d-1)} R^{d-1} \sim \al \phi_0^2
|\tau|^{2 \beta} R^{d + 2}.
\end{eqnarray}
Rescaling the order parameter and length appropriately and introducing
the {\em renormalized} coupling constant
\begin{eqnarray} \label{ep:ren}
\ep^2 = \al \phi_0^2 \xi_0^{3} |\tau|^{2 \beta - \nu (d + 2)},
\end{eqnarray}
we can still write down the interfacial free energy of the system in
the form of Eq.~(\ref{F:inter}), where, as usual, we dropped the
primes and neglected an overall constant factor. Caution, however, is
necessary here in considering the screening effects. As was noted
earlier, the constant $\kappa$ appearing in the mean-field definition
of the screened long-range interaction $G_\ep (x - x')$ is related to
the coefficient of linear response for the local theory. When the
critical phenomena fluctuations are taken into account, the value of
$\kappa$ can be calculated via the linear response function $\chi =
\chi_0 |\tau|^{-\gamma}$ that relates the unscaled values of
$\phi_\mathrm{sm}$ and $\psi$ (below $T_c$), see Eqs.~(\ref{lc}) and
(\ref{psi}). After an appropriate rescaling and using the definition
of $\ep$ from Eq.~(\ref{ep:ren}), we obtain that
\begin{eqnarray}
 \label{kappa} \kappa^2 = { \chi_0 \over \phi_0^2 \xi_0},
\end{eqnarray}
which is a constant of order one, independent of temperature. In
writing Eq. (\ref{kappa}) we used the scaling relation $2 \beta +
\gamma - d \nu = 0$ between the critical exponents \cite{landau5}.

Thus, we can renormalize the parameter $\ep$ and redefine the
parameter $\kappa$ to obtain once again the free energy of a domain
pattern in the form of Eq. (\ref{F:inter}) even in the case of the
system locally experiencing critical phenomena fluctuations. Let us
point out that the free energy of a domain pattern of size of order $R
\gg \xi$ will be much greater than $k_B T_c$ (in the unscaled units),
see Eq.~(\ref{F:inter}), since $\sigma \xi^{d-1} / k_B T_c \sim 1$
\cite{landau5}. This leads us to the conclusion that in a strongly
fluctuating system the domain patterns should be essentially described
by the interfacial mean-field theory, and all the properties of the
domains in the fluctuating system will be equivalent to those of the
domains in the mean-field systems described by Eq.~(\ref{F}), provided
that one uses the renormalization given by Eqs. (\ref{ep:ren}) and
(\ref{kappa}). Thus, the universality discussed earlier for the
mean-field model should in fact extend to all systems near the local
critical temperature as long as the coupling constant $\al$ of the
long-range interaction is small enough. Note that in this situation
Eq.~(\ref{F}) may be used as a phase-field model representation for
the free energy of the domain patterns \cite{m1:prl97}.

In the renormalization of the main parameters of the system we made an
assumption that the size of the domains must be much greater than the
correlation length $\xi$. According to Eq. (\ref{R:gen}), this
condition is satisfied as long as $\ep \ll 1$.  In view of
Eq. (\ref{ep:ren}), this is the case when the reduced temperature
$\tau$ is much lower than $\tau = -\tau_c$, where
\begin{eqnarray}
\tau_c \sim \al^{1 / [\nu (d + 2) - 2 \beta ]},
\end{eqnarray}
at which $\ep \sim 1$. When the temperature is decreased below
$-\tau_c$, the long-range interaction becomes progressively more and
more relevant at long distances, until for $\tau \ll -\tau_c$ (what
means $\ep \ll 1$) long-lived domain structures with the free energy
cost of each domain $\Delta F / k_B T_c \gg 1$ will start to form.  On
the other hand, for $\tau \gg +\tau_c$ the long-range coupling, which
scales as $\al \xi^2$, will be much smaller than the effective local
coupling, which is of order $\chi^{-1} \sim |\tau|^\gamma$, so one
will observe only the critical phenomena fluctuations above
$\tau_c$. It is a question whether there is a microphase separation
transition from the homogeneous to the patterned phase (which is
analogous to the freezing transition in liquids) or there is a smooth
crossover from one to another in a strongly fluctuating system. It is
clear, however, that the uniform phase must be thermodynamically
unstable when $\tau \lesssim - \tau_c$. At the same time, at $\tau
\sim -\tau_c$ the fluctuations are strong, so one can envisage the
system as a collection of domains that randomly appear and disappear
in different locations and move about as particles in a dense
liquid. In any case, there must exist a narrow transition region
$-\tau_c \lesssim \tau \lesssim \tau_c$, upon going through which the
phase should change from uniform to patterned.

\section{Properties of the domain patterns} 

\subsection{Equations for stationary patterns} \label{s:stat}

The stationary patterns in the mean-field model introduced in
Sec. \ref{s:mf} must satisfy the Euler-Lagrange equation obtained from
Eq.~(\ref{F}): 
\begin{eqnarray} \label{el}
\nabla^2 \phi + \phi - \phi^3 && \nonumber \\ && \hspace{-2cm} + \mu -
\ep^2 \int d^d x' G(x - x') (\phi(x') - \phb) = 0,
\end{eqnarray}
where $\mu$ is the chemical potential.  Formally, this
integrodifferential equation can be rewritten as a pair of stationary
reaction-diffusion equations of activator-inhibitor type
\cite{ohta90,m1:prl97,petrich94,goldstein96}. Indeed, if the last term
in Eq.~(\ref{el}) is denoted by $\psi$, this equation can be rewritten
as
\begin{eqnarray}
\nabla^2 \phi + \phi - \phi^3 - \psi & = & 0, \label{act} \\ \nabla^2
\psi + \ep^2 (\phi - \phb) & = & 0, ~~~\langle \psi \rangle = -\mu,
\label{inh}
\end{eqnarray}
where $\langle \cdot \rangle$ denote averaging over the system's
volume.
 
Reaction-diffusion equations of the type of Eqs.~(\ref{act}) and
(\ref{inh}) have been studied by many authors (see, for example,
\cite{ko:book,ohta89,mo1:pre96,goldstein96,hagberg:prl94,nishiura98}). In
the limit $\ep \rightarrow 0$ their solutions can be treated by the
methods of singular perturbation theory (matched asymptotics)
\cite{mo1:pre96,vasileva,ko:book,fife}. According to singular
perturbation theory, the solution $\phi$ of Eq.~(\ref{act}) can be
broken up into the inner and outer solutions. The inner solution
varies on the length scale of order 1 and describes the variation of
the order parameter in the vicinity of the domain interfaces, while
the outer solution varies on the length scale $R$ of the order of the
characteristic size of the domains and describes the variation of the
order parameter away from the interfaces. Similarly, the solution
$\psi$ of Eq.~(\ref{inh}) will vary on the length scale $R$.

Since the variable $\psi$ varies slowly on the inner scale, it can be
considered as constant in the interface. Since the curvature of the
domain wall is also much smaller than the domain wall width, to the
leading order we can write Eq.  (\ref{act}) in the vicinity of the
interface as
\begin{eqnarray} \label{qqq}
 \label{qq} {\partial^2 \phi \over \partial \rho^2} - 2 H {\partial
\phi \over \partial \rho} + \phi - \phi^3 - \psi_i = 0,
\end{eqnarray}
where $\rho$ is the distance from a given point to the interface,
which is positive if the point is inside the positive domain and
negative otherwise, $H = \frac{1}{2} (k_1 + k_2)$ is the mean
curvature of the interface (positive if the positive domain is convex,
$k_1$ and $k_2$ are the principal curvatures), and $\psi_i$ is the
value of $\psi$ on the interface. In the following, we will write all
the formulas in the three-dimensional case, in two or one dimensions
one has to set one or two principal curvatures of the interface,
respectively, to zero.

Equation (\ref{qqq}) can be solved exactly, its solution has the form
$\phi(\rho) = a \tanh b \rho + c$, where $a$, $b$, and $c$ are certain
constants. This solution exists only when $2 H - \frac{16}{9} H^3 = -
3 \psi_i/\sqrt{2}$. Since in the domain pattern $H \ll 1$, it is
sufficient to linearize this equation with respect to $H$, so we
obtain (see also \cite{m:pre96})
\begin{eqnarray}
 \sigma_0 \label{v} H = - \psi_i,
\end{eqnarray}
where $\sigma_0$ is given by Eq.~(\ref{interf}). This, in turn,
implies that $\psi \ll 1$ in order for a pattern to be
stationary. Note that to the leading order $\phi(\rho)$ in the
interface is given by Eq.~(\ref{sh}).

Away from the interfaces (on the outer scale) $\phi$ varies slowly, so
one can neglect the gradient square term in Eq. (\ref{act}). Then,
according to Eq. (\ref{act}), we have $\phi - \phi^3 = \psi$. Since we
must have $\psi \ll 1$, this equation can be linearized with respect
to $\phi$ around $\phi = \pm 1$ in the positive and negative domains,
respectively.  So, one obtains that to the leading order $\phi = \pm
1- \kappa^2 \psi$ in the outer regions. Here, as in Sec. \ref{s:int},
we have $\kappa^2 = \frac{1}{2}$. If we substitute this expression
into Eq. (\ref{inh}), we will obtain precisely the same equation for
$\psi$ as Eq. (\ref{psiscreened}), with $\phi_\mathrm{sh} = \pm 1$.

The solution of Eq. (\ref{psiscreened}) can be written as an integral
over the domain interfaces (Appendix \ref{a:free}, see also
  \cite{goldstein96,m:pre96}):
\begin{eqnarray}
 \label{psicont} \psi = - {1 + \phb \over \kappa^2} +
 \frac{2}{\kappa^2} \oint dS' \{ \hat{n}' \cdot \vec{\nabla}' (G_\ep -
 G)\},
\end{eqnarray}
where $\vec{\nabla}'$ is the gradient with respect to $x'$. Combining
this equation with Eq. (\ref{v}), we obtain the following equation for
the locations of the interfaces
\begin{eqnarray}
 \label{v:ma} \sigma_0 H = {1 + \phb \over \kappa^2} - {2 \over
 \kappa^2} \oint dS' \{ \hat{n}' \cdot \vec{\nabla}' (G_\ep - G) \}.
\end{eqnarray}
This equation can be further simplified if the distance between the
points on the interface is much smaller than $\ep^{-1}$. In this case
one can expand $G_\ep $ in Eq. (\ref{v:ma}) in $\ep \kappa |x - x'|$
and retain the terms up to the second order. It is easy to see that
only the terms of the second and higher orders of the expansion of
$G_\ep$ in $\ep$ will give non-trivial contributions to the right-hand
side of Eq.~(\ref{v:ma}). Also, in view of the approximations used to
derive Eq.~(\ref{v:ma}), this equation is valid when the
characteristic size $R$ of the domains satisfies $1 \ll R \ll
\ep^{-1}$.

Equation (\ref{v:ma}) describes the pressure balance across the
interface. Indeed, the term in the left-hand side of this equation is
the Laplace law, the first term in the right-hand side gives the bulk
pressure, and the second one gives the nonlocal contribution to
pressure due to the interaction of the domain walls with each other.

Let us emphasize that Eq.~(\ref{v:ma}) can also be straightforwardly
obtained by computing the first variation of the interfacial free
energy given by Eq.~(\ref{F:inter}) (see Appendix
\ref{a:free}). Therefore, this equation also remains valid in the
fluctuating system considered in Sec. \ref{s:renorm}. Also, note that
since the solutions of Eq.~(\ref{el}) in the form of stationary domain
patterns can be written in the form of Eq.~(\ref{shsm}) for $\ep \ll
1$, Eq.~(\ref{F:inter}) gives the asymptotic expression for the free
energy of these patterns.

\subsection{Deformations of the domain interfaces} \label{s:fluct}

The solutions of the Euler-Lagrange equation given by Eq.~(\ref{el})
are critical points of the free energy functional from
Eq.~(\ref{F}). Similarly, the solutions of Eq.~(\ref{v:ma}) are
critical points of the interfacial free energy from
Eq.~(\ref{F:inter}) and correspond to the solutions of Eq.~(\ref{el})
in the limit $\ep \rightarrow 0$. Both these solutions define
(generally, metastable) stationary patterns. The question, however,
arises as to when these patterns are thermodynamically {\em
stable}. Since, apart from the nucleation phenomena discussed in
Sec. \ref{s:nucl}, the effect of thermal fluctuations is small in both
cases, the thermodynamic stability of the patterns is determined by
the second variation of the free energy functional. Thus, the
thermodynamically stable stationary patterns will be local minimizers
of the free energy.

The problem of finding the second variation of the functional in
Eq.~(\ref{F}) reduces to the calculation of the spectrum of
linearization of Eq.~(\ref{el}). It is not difficult to see that it is
equivalent to the problem of linear stability of stationary patterns
in systems obeying gradient descent dynamics. Such a stability
analysis in the context of general reaction-diffusion systems of
activator-inhibitor type was performed in \cite{mo1:pre96}. Here,
instead of analyzing the second variation of $F$ from Eq.~(\ref{F}),
we will use the interfacial free energy from Eq.~(\ref{F:inter}) for
finding the spectrum of the fluctuations of the pattern's
interfaces. These are, in turn, the lowest-lying modes of the spectrum
and, therefore, cost the least free energy in Eq.~(\ref{F}). Both
these approaches give the same results in the limit of small $\ep$.

Let us now proceed with the calculation of the second variation of the
interfacial free energy from Eq.~(\ref{F:inter}). A small perturbation
of the domain shape means a slight shift of the interface in the
normal direction by $\rho(x)$, where $x$ denotes a point on the
interface. In terms of $\rho(x)$, the second variation of the free
energy from Eq.~(\ref{F:inter}) is (see Appendix \ref{a:free})
\begin{eqnarray}
\delta^2 F = \sigma_0 \oint d S \left\{ |\nabla_\perp \rho|^2 + 2 K
\rho^2 - 4 H^2 \rho^2 \right\} \nonumber \\ + {4 \over \kappa^2} \oint
d S ~\rho^2 (\hat n \cdot \vec\nabla) \oint d S' \{ \hat n' \cdot
\vec\nabla' (G_\ep - G) \} \nonumber \\ + 4 \ep^2 \oint d S \oint d S'
G_\ep(x - x') \rho(x) \rho(x'), \label{matr}
\end{eqnarray}
where $\nabla'$ is the gradient in $x'$, $K = k_1 k_2$ is the Gaussian
curvature at a given point on the unperturbed interface,
$\nabla_\perp$ is the gradient along the interface, and the
integration is over the unperturbed interfaces. 

Different terms in the integrand of Eq. (\ref{matr}) represent
competing tendencies that stabilize or destabilize the patterns. The
$\sigma_0 |\nabla_\perp \rho|^2$ term coming from the surface tension
penalizes the distortions of the interfaces; the term involving the
curvatures $2 \sigma_0 (K - 2 H^2)\rho^2 = - \sigma_0 ( k_1^2 + k_2^2)
\rho^2 \leq 0$, is a destabilizing term coming from the curvature of
the interface; the term from the second line in Eq.~(\ref{matr}) can
be rewritten as $2 (\hat n \cdot \vec\nabla \psi) \rho^2$, where
$\psi$ is given by Eq.~(\ref{psicont}) (see Appendix \ref{a:free}),
and represents the change in the free energy due to the motion of the
interface in the fixed effective field $\psi$, this term should be
destabilizing also since we would generally expect the gradient of
$\psi$ to be directed inward at the interface; and the last term is a
stabilizing action of the long-range interaction.

To gauge the relative strengths of these terms and determine whether a
pattern is stable, we need to solve the following eigenvalue problem
obtained from Eq.~(\ref{matr}):
\begin{eqnarray} \nonumber \label{lin}
L \rho & = & \lambda \rho, \mathrm{~~where} \nonumber \\ L \rho & = &
-\sigma_0 \nabla_\perp^2 \rho + 2 \sigma_0 K \rho - 4 \sigma_0 H^2
\rho \\ && + 2 (\hat n \cdot \vec\nabla \psi) \rho + 4 \ep^2 \oint dS'
G_\ep(x - x') \rho(x'). \nonumber
\end{eqnarray}
The spectrum of the operator $L$ for a stable pattern should not
contain any negative eigenvalues. We will analyze the spectrum of $L$
for simple geometries below (see also
\cite{mo1:pre96,mo2:pre96,m1:prl97,ohta89,goldstein96,yeung94}). Now,
however, let us discuss some general properties of $\delta^2 F$ in
Eq.~(\ref{matr}). It is easy to see from Eq.~(\ref{matr}) that a
typical size $R$ in a {\em stable} stationary pattern must have the
same scaling as that in Eq.~(\ref{R:gen}) (this point was first argued
in \cite{mo1:pre96,mo2:pre96,m1:prl97} based on the stability analysis
of the localized and periodic patterns). Indeed, suppose a pattern is
made of a collection of droplets of size and distance between each
other of order $R$ (here for definiteness we will consider
three-dimensional patterns). Let us first assume that the droplets are
too small, so $R \ll \ep^{-2/3}$. Consider a fluctuation that
increases uniformly the volume of one droplet while decreasing the
volume of another next to it, so that the net volume change is zero
(repumping, see \cite{ko:book,m1:prl97}). Then, if $R \ll \ep^{-2/3}$,
the stabilizing contribution $\sim \ep^2 R^3$ from the last term in
Eq.~(\ref{matr}) is negligible compared to the destabilizing
contribution from the curvature terms $\sim 1$, while the
$|\nabla_\perp \rho|^2$ term is identically zero. Therefore, such a
fluctuation will lead to the free energy decrease.

Now, suppose that the droplets are too big, so $R \gg \ep^{-2/3}$. Let
us now perturb the interface of one droplet in a localized fashion in
the region of size $\ell \ll R$, once again, maintaining the overall
volume the same (distortion, \cite{ko:book,mo1:pre96,m1:prl97}). Then
the last term in Eq.~(\ref{matr}), which is $\sim \ep^2 \ell^3$ will
once again be negligibly small compared to the term from the second
line of Eq.~(\ref{matr}), which is $\sim \ep^2 \ell^2 R$. On the other
hand, the gradient square term in Eq.~(\ref{matr}), which is of order
$\sim 1$ will not be able to compensate that contribution, if
$\ep^{-1} R^{-1/2} \ll \ell \ll R$. Such $\ell$ can always be found
when $R \gg \ep^{-2/3}$, so this kind of a fluctuation will lower the
free energy, too. Note that for $R \sim \ep^{-1}$ this instability
result was also obtained by Nishiura and Suzuki \cite{nishiura98}.

The arguments above lead to an important conclusion that (perhaps,
apart from some logarithmic factors, see below) the stable stationary
patterns must obey the equilibrium scaling from Eq.~(\ref{R:gen}),
which was obtained on global energetic grounds. In other words, not
only the global minimizers of the free energy \cite{choksi01}, but all
{\em local} minimizers must generally obey this scaling. Note,
however, that these arguments do not apply in one dimension (see also
\cite{ko:book,mo1:pre96,ohta89,onishi99}). Similarly, the equilibrium
scaling from Eq.~(\ref{R:gen}) is not necessarily obeyed by all
stationary patterns, see for example, Sec. \ref{s:sol} and
\ref{s:hexlam}, contrary to the statement of \cite{ nishiura95}.

\section{Equilibrium patterns and morphological instabilities} 

Let us now use the tools developed in the preceding sections to
analyze the stationary domain patterns with simple geometries, such as
localized and periodic patterns. In this paper, we will limit
ourselves to studying two dimensional patterns. Qualitatively the same
results are expected for the more experimentally relevant
three-dimensional patterns. Note, however, that the one-dimensional
case is qualitatively different from two and three dimensions (see
\cite{onishi99,ko:book,mo1:pre96,m1:pre97}).

Since our system possesses a symmetry $\phi \rightarrow - \phi$, we
can only consider the properties of the positive domains immersed in
the negative background. This means that we only need to study the
region of the system's parameters in which $\phb < 0$.

\subsection{Solitary patterns} \label{s:sol}

We begin with the study of the simplest possible domain patterns:
solitary patterns. In two dimensions we will consider spots, stripes,
and the annuli.

\subsubsection{Spot} 

\begin{figure}
\centerline{\psfig{figure=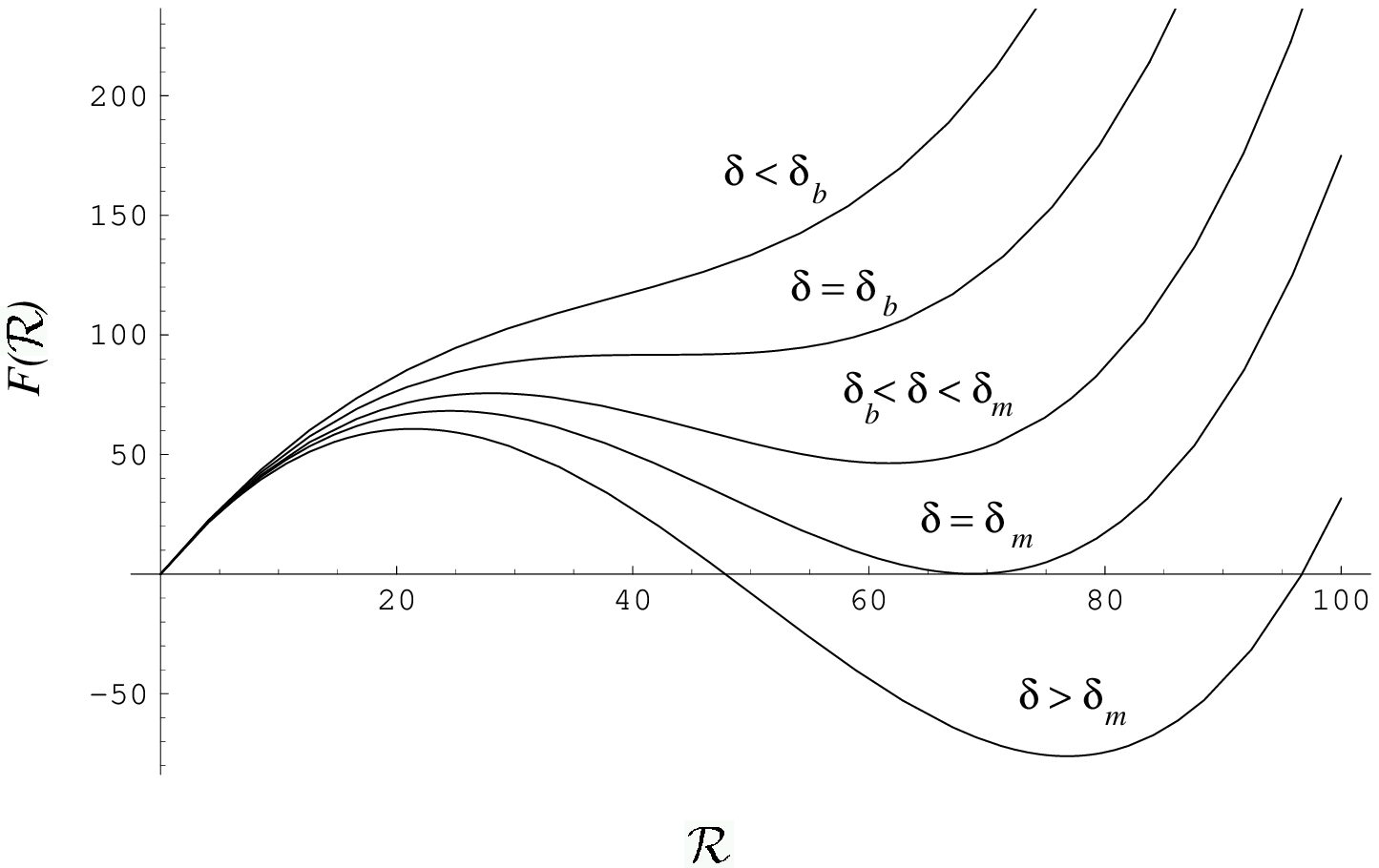,width=3in}} 
\caption{The free energy of a spot for different values of $\phb$. The
plot of $F(\rs)$ from Eq. (\protect\ref{Rspot}) with $\ep = 0.001$,
$\sigma_0 = 2 \sqrt{2}/3$, and $\kappa = 1/\sqrt{2}$. For these values
of the parameters $\delta_b = 0.0088$ and $\delta_m = 0.0186$. }
\label{f:rspot}
\end{figure}

Let us first look at a spots: a small positive circular domain in the
negative background. If the radius of the spot is much smaller than
the screening length $\ep^{-1}$, the interaction potential $G_\ep(x -
x')$ in Eq. (\ref{F:inter}) can be expanded in $\ep$. Retaining the
terms up to $\ep^2$, after a straightforward calculation we obtain
that the free energy of a spot of radius $\rs$ is asymptotically
\begin{eqnarray}
 \label{Rspot} F(\rs) & = & 2 \pi \sigma_0 \rs - {2 \pi \rs^2 \delta
\over \kappa^2} \nonumber \\ && -\pi \ep^2 \rs^4 \left[ \ln \left( {1
\over 2} \ep \kappa \rs \right) + \gamma - \frac{1}{4} \right],
\end{eqnarray}
where $\gamma \simeq 0.5772$ is the Euler constant and 
\begin{eqnarray} \label{delta}
\delta = 1 + \phb
\end{eqnarray}
measures the degree of metastability of the homogeneous phase. The
free energy of the spot given by Eq.  (\ref{Rspot}) for a particular
set of parameters is shown in Fig. \ref{f:rspot}.

Let us now analyze Eq (\ref{Rspot}). First of all, when $\delta < 0$,
the free energy is a monotonically increasing function of $\rs$. When
the value of $\delta$ is increased, at $\delta = \delta_b \ll 1$ a
minimum and a maximum of the free energy appear (see
Fig. \ref{f:rspot}). These correspond to the radially stable and
unstable spot solutions, with the radii $\rs = \rs_s$ and $\rs =
\rs_n$, respectively. At $\delta = \delta_b$ we have $\rs_s = \rs_n =
\rs_\mathrm{min}$. Asymptotically
\begin{eqnarray}
\delta_b = \left( { 3 \over 4} \ep \kappa^3 \sigma_0 \ln^{1/2}
\ep^{-1} \right)^{2/3} \sim \ep^{2/3} \ln^{1/3} \ep^{-1},
\label{deltab} \\ \rs_\mathrm{min} = \left( { 3 \sigma_0 \over 4 \ep^2
\ln \ep^{-1} } \right)^{1/3} \sim \ep^{-2/3} \ln^{-1/3}
\ep^{-1}. \label{Rmin}
\end{eqnarray}
This formula agrees up to the logarithmic factor with
Eq.~(\ref{R:gen}). These logarithmic factors are a specific feature of
the two-dimensional patterns, they are absent in three dimensions
\cite{mo1:pre96}.

When the value of $\delta$ is increased beyond $\delta_b$, the radius
$\rs_s$ grows, while the radius $\rs_n$ shrinks. At some value of
$\delta = \delta_m \ll 1$ at which $\rs_s = \rs_m$, the free energy of
the spot becomes negative, making the spot thermodynamically more
favorable than the homogeneous phase. Once again, asymptotically,
\begin{eqnarray}
\delta_m = \frac{1}{2} \left( 3 \ep \kappa^3 \sigma_0 \ln^{1/2}
\ep^{-1} \right)^{2/3} \sim \ep^{2/3} \ln^{1/3} \ep^{-1},
\label{deltam} \\ \rs_m = \left( { 3 \sigma_0 \over \ep^2 \ln \ep^{-1}
} \right)^{1/3} \sim \ep^{-2/3} \ln^{-1/3} \ep^{-1}. \label{Rm}
\end{eqnarray}
Comparing Eqs.~(\ref{deltab}) -- (\ref{Rm}), we see that $\delta_m =
2^{1/3} \delta_b$, and $\rs_m = 2^{2/3} \rs_\mathrm{min}$.

For $\delta \gg \delta_b$ the radii $\rs_s$ and $\rs_n$ become
asymptotically
\begin{eqnarray} \label{rsd}
\rs_s = \left( { 3 \delta \over \ep^2 \kappa^2 \ln \ep^{-1} }
\right)^{1/2}, ~~~\rs_n = {\sigma \kappa^2 \over 2 \delta}.
\end{eqnarray}
This means that for $\delta \gg \ep^{2/3} \ln \ep^{-1}$, the radius
$\rs_s$ goes beyond the equilibrium scaling of Eq.~(\ref{R:gen}). This
is an indication of a morphological instability studied in
Sec. \ref{s:morsol}.

\subsubsection{Annulus}

Let us now analyze the pattern in the form of a thin annulus, which
has the radius $\rs$ and thickness $\ls \ll \rs$. Calculating the free
energy of such a pattern from Eq.~(\ref{F:int}), we obtain
\begin{eqnarray} \label{F:ann}
F(\rs, \ls) & = & 4 \pi \sigma_0 \rs - {4 \pi \delta \over \kappa^2}
\rs \ls \nonumber \\ & + & 4 \pi \ep^2 \rs^2 \ls^2 I_0(\ep \kappa \rs)
K_0(\ep \kappa \rs),
\end{eqnarray}
where $I_0(x)$ is the modified Bessel function. Minimizing this
expression with respect to $\ls$, we obtain that the value of $\ls =
\ls_a$ in equilibrium is related to $\rs$ as follows
\begin{eqnarray} \label{La}
\ls_a = {\delta \over 2 \ep^2 \kappa^2 \rs I_0(\ep \kappa \rs) K_0(\ep
\kappa \rs)}.
\end{eqnarray}
Substituting this expression into Eq.~(\ref{F:ann}), we then study the
critical points of $F$ with respect to $\rs$.

The analysis of Eqs.~(\ref{F:ann}) and (\ref{La}) shows that for
$\delta \ll \ep^{1/2}$ there exists a single minimum of the free
energy, corresponding to an annulus solution, whose radius and width
are asymptotically
\begin{eqnarray} \label{RaLa}
\rs_a = {\delta^2 \over 4 \sigma_0 \ep^2 \kappa^4 \ln^2 (\ep
\delta^{-2}) }, ~\ls_a = {2 \sigma_0 \kappa^2 \over \delta} \ln (\ep
\delta^{-2}).
\end{eqnarray}
One can see that the condition $\ls_a \ll \rs_a$ used in the
derivation of Eq.~(\ref{RaLa}) is satisfied as long as $\delta \gg
\ep^{2/3} \ln \ep^{-1}$.

\begin{figure}[t]
\centerline{\psfig{figure=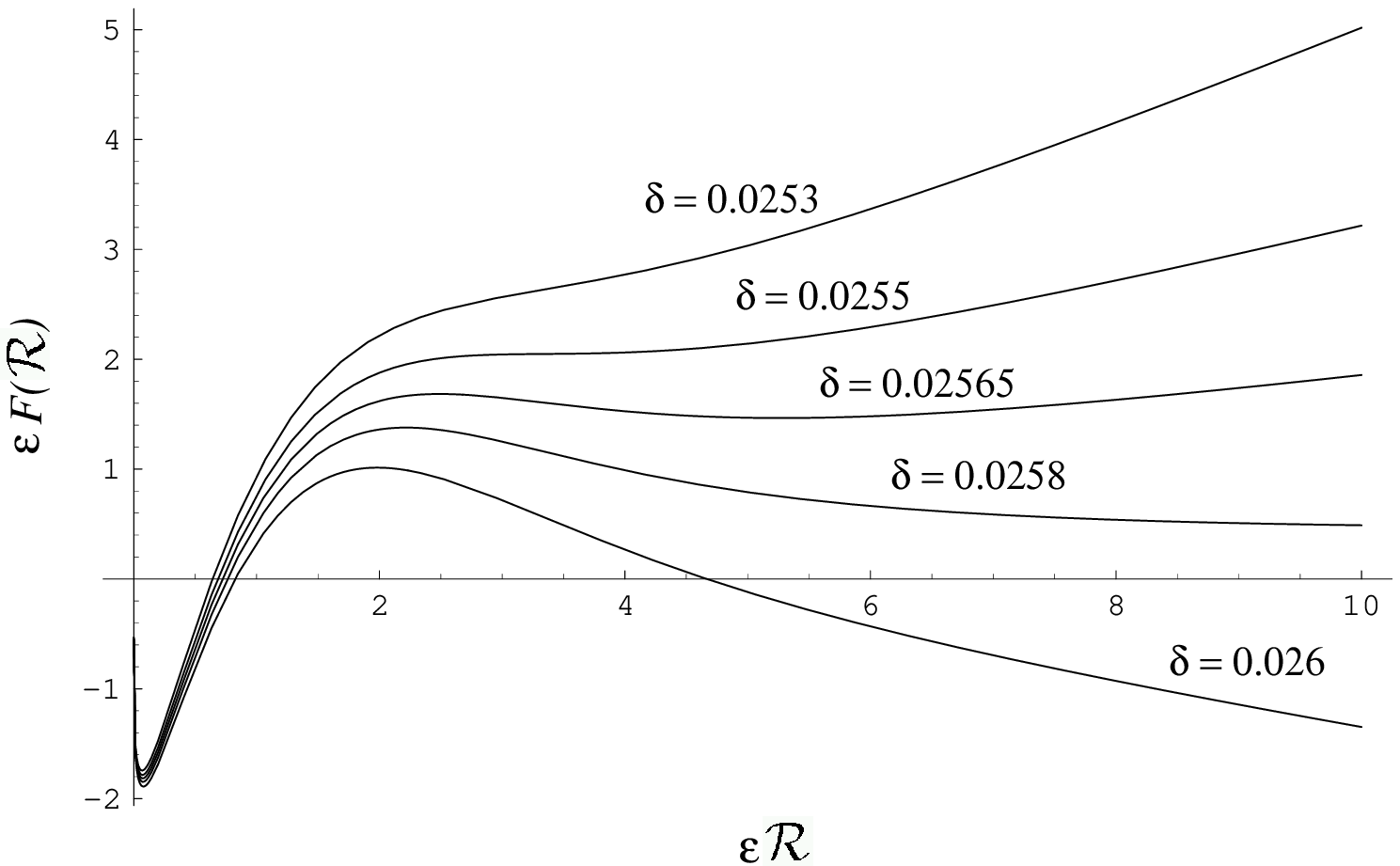,width=3in}}
\caption{The free energy of an annulus as a function of $\rs$ for $\ls
= \ls_a$ given by Eq.~(\ref{La}), obtained from Eq.~(\ref{F:ann}) with
$\ep = 0.001$, $\sigma_0 = 2 \sqrt{2} / 3$, and $\kappa = 1 /
\sqrt{2}$.}
\label{f:ann}
\end{figure}

According to Eq.~(\ref{RaLa}), when $\delta \sim \ep^{1/2}$, we have
$\rs_a \sim \ep^{-1}$, so screening effects become important. The
analysis of Eq.~(\ref{F:ann}) shows that at some critical value of
$\delta \sim \ep^{1/2}$, a new minimum and a maximum of $F(\rs)$
appear ($\delta = 0.0255$ in Fig. \ref{f:ann}).
At a slightly higher value of $\delta \sim \ep^{1/2}$, the second
minimum of the free energy disappears ($\delta = 0.0258$ in
Fig. \ref{f:ann}). The value of $\delta = \delta_\perp$ at which this
happens can be easily calculated, see Eq.~(\ref{deltaperp}). 

\subsubsection{Stripe}

Let us now determine the equilibrium parameters of a quasi
one-dimensional domain pattern --- stripe. A stripe of width $\ls_s$
can be considered as a limit of an annulus as $\rs_a \rightarrow
\infty$. Using Eq.~(\ref{F:ann}), we obtain that the free energy of a
stripe of length $\ls$ is
\begin{eqnarray} \label{F:stripe}
F = \left( 2 \sigma_0 - {\delta^2 \over \ep \kappa^3} \right) \ls.
\end{eqnarray}
The term in the brackets characterizes the rigidity of a stripe. As
can be seen from Eq.~(\ref{F:stripe}), this rigidity becomes negative
at a critical value of $\delta = \delta_\perp$, where
\begin{eqnarray}
\label{deltaperp}
\delta_\perp = (2 \sigma_0 \kappa^3 \ep)^{1/2}.
\end{eqnarray}
At $\delta > \delta_\perp$ the stripe becomes unstable with respect to
wriggling (see Sec. \ref{s:morsol}).

Taking the limit $\rs_a \rightarrow \infty$ in Eq.~(\ref{La}), we
obtain
\begin{eqnarray} \label{lsstripe}
 \label{ls} \ls_s = { \delta \over \ep \kappa}.
\end{eqnarray}
Note that the stripe solutions of Eq.~(\ref{el}) exist only when
$\ls_s \gtrsim \ln \ep^{-1}$, so to have a solution we must have
$\delta \gtrsim \ep \ln \ep^{-1}$ \cite{ko:book,mo1:pre96}. Also, the
width of a stripe is limited by $\ls_s \sim \ep^{-1}$, in which case
our linearization approximation to Eq.~(\ref{euler:sm}) breaks down
(see also   \cite{ko:book}). According to Eq.~(\ref{lsstripe}), the
region of existence of stripes is wider than that of spots. Also, note
that for $\delta \sim \delta_b$ (when the spot is stable), the width
of the stationary stripe $\ls_s \sim \ep^{-1/3} \ll \rs_s$. This
deviation from the equilibrium length scale given by Eq.~(\ref{R:gen})
is essentially related to the one-dimensional nature of the stripe,
for which the curvature effects are absent.

\subsection{Hexagonal and lamellar patterns} \label{s:hexlam}

When the spots or stripes are introduced into the system, the basic
interaction between them is repulsion [see Eq. (\ref{F:int})]. In an
equilibrium configuration, the domains will therefore go as far apart
from each other as possible. If in the end the distance between them
is greater than the screening length $\ep^{-1}$, essentially they will
not interact, so their behavior will be that of the solitary patterns
discussed in Sec. \ref{s:sol}. The situation changes, however, when
there are so many domains in the system that even in the close-packed
arrangement the distance between them becomes less than the screening
length. This is in fact a generic situation that is realized whenever
the value of $\phb$ is not close to $-1$ (or $\delta \sim 1$). In this
case the domains will strongly interact with each other, arranging
themselves into a {\em multidomain} pattern, so in order to decrease
the energy of the long-range repulsion, the domains not only adjust
their positions, but also their geometric characteristics.

Let us consider the simplest of the multidomain patterns in two
dimensions, namely, the periodic hexagonal and lamellar patterns. The
equilibrium characteristics for several major types of periodic
patterns described by Eq.~(\ref{F}) in the limit $\ep \rightarrow 0$
were found by Ohta and Kawasaki \cite{ohta86}. They carried out a
rather involved calculation of the free energy using the Ewald
summation method. Their results can be obtained by the simpler,
although approximate, Wigner-Seitz method \cite{ziman}. Consider a
hexagonal pattern made up of circular domains, for example. In such a
pattern $\psi$ will satisfy Eq.~(\ref{psiscreened}) with no flux
boundary conditions on the boundaries of the hexagonal Wigner-Seitz
cell. Instead of solving this problem, let us consider a single domain
inside a circular cell whose area is equal to the area of the
Wigner-Seitz cell (a similar approach was used in
  \cite{m1:pre97}). Then Eq. (\ref{psiscreened}) with no flux
boundary conditions can be easily solved. Furthermore, to the leading
order in $\ep$ the screening term $\ep^2 \kappa^2 \psi$ in
Eq.~(\ref{psiscreened}) can be neglected, if the period of the pattern
$\ls_p \ll \ep^{-1}$. This solution can be used to calculate the
contribution from the long-range interaction to the free energy by
substituting it to Eq.~(\ref{longcontr}). Note that the
one-dimensional analog of this method is exact, so it can also be used
to calculate the free energy of the lamellar pattern.

Let the positive domains in a hexagonal pattern have radius $\rs_s$
and period $\ls_p \ll \ep^{-1}$. It is convenient to introduce the
fraction $f$ of the total area of the system occupied by positive
domains. The condition of Eq.~(\ref{conserv}) implies that this
fraction is related to $\phb$ as
\begin{eqnarray}
 \label{fraction}
 f = { 1 + \phb \over 2},
\end{eqnarray}
since inside the domains (away from the interfaces) $\phi \simeq \pm
1$ [see the discussion after Eq.~(\ref{Gsc})]. In the hexagonal
pattern $\rs_s$ and $\ls_p$ are related via
\begin{eqnarray}
 \label{rslp2}
 \rs_s = 3^{1/4} \ls_p \left( {1 + \phb \over 4 \pi} \right)^{1/2}.
\end{eqnarray}
From this equation one can see that when $\phb$ is not close to $-1$,
the values of $\rs_s$ and $\ls_p$ are comparable. Note that in reality
Eq. (\ref{rslp2}) is approximate, since generally the domains forming
a hexagonal pattern are not ideally round. However, according to the
numerical simulations, the deviations from the circular shape are very
small when $\phb < 0$, so one can safely assume the domains to be
ideally circular all the way up to $\phb = 0$.

In Eq. (\ref{rslp2}) the period of the pattern has not been specified.
In fact, an infinite set of solutions in the form of hexagonal
patterns with different periods exists for $-1 < \phb < 0$
(asymptotically). All these solutions locally minimize the free energy
of the system. However, among all hexagonal patterns there is a
pattern with a particular period $\ls_p^*$ for which the value of the
free energy is the lowest. It is clear that if the asymmetry between
the positive and negative domains is strong, the domains will tend to
form a close-packed structure, so for $\delta \ll 1$ in $d = 2$ this
pattern is expected to be the global minimizer of the free energy.

Using the Wigner-Seitz method, we find that the period $\ls_p^*$ of
the hexagonal pattern with the lowest free energy is (Appendix
\ref{a:hexlam})
\begin{eqnarray}
 \label{lp2eq} \ls_p^* = \ep^{-2/3} \left( {2 \pi \over f \sqrt{3} }
 \right)^{1/2} \left( {2 \sigma_0 \over f - 1 - \ln f }
 \right)^{1/3}.
\end{eqnarray}
As ought to be expected, $\rs_s \sim \ls_p^* \sim \ep^{-2/3}$. It is
interesting to note that this result agrees with the exact calculation
of Ohta and Kawasaki in   \cite{ohta86} to within $0.1\%$ for all
$f < 0.5$.

Similarly, in the case of the lamellar pattern the period $\ls_p$ and
width $\ls_s$ of the stripe are related as
\begin{eqnarray}
 \label{lslam}
 \ls_s = {1 + \phb \over 2} \ls_p.
\end{eqnarray}
A calculation analogous to the case of the hexagonal pattern shows
that the period of the lamellar pattern that has the lowest free
energy is given by (Appendix \ref{a:hexlam})
\begin{eqnarray}
 \label{lp1eq} \ls_p^* = \ep^{-2/3} \left( {6 \sigma_0 \over f^2 (1 -
 f)^2 } \right)^{1/3}.
\end{eqnarray}
This result agrees with that of
\cite{onishi99,yeung94,ohta86,muller93,ren00}. Let us point out that
in one dimension Ren and Wei proved that the lamellar patterns of
arbitrary periods are the only local minimizers of the free energy
\cite{ren00}.

A comparison of the free energies per unit area of the lowest free
energy hexagonal and lamellar patterns shows that the lamellar pattern
has lower free energy when $f > 0.35$ \cite{ohta86}. For $0 < f <
0.35$ the hexagonal pattern has the lowest free energy in two
dimensions.

\subsection{Morphological instabilities of solitary patterns}
\label{s:morsol} 

An important feature of patterns in systems with competing
interactions is the fact that under certain conditions they can
undergo {\em morphological instabilities} which lead to the
distortions of their shapes and transitions between them
\cite{seul95}. In reaction-diffusion systems these instabilities have
been analyzed in
\cite{ko:book,mo1:pre96,mo2:pre96,m1:prl97,ohta89,hagberg:prl94,
goldstein96,pismen94}.

Apart from the arguments of Sec. \ref{s:fluct}, the physical reason
for the existence of morphological instabilities is the fact that the
energy of the long-range interaction increases faster than the area of
the domain as its size gets bigger. Therefore, at some critical size
it may become energetically favorable for the domain to split into two
domains of smaller size or significantly change its shape. It is
interesting to note that such an instability was first analyzed by
Lord Rayleigh back in 1882 \cite{rayleigh1882}.

To investigate the morphological instabilities of the domain patterns
in systems with long-range Coulomb interaction, we start by looking at
the simplest possible patterns: spots and stripes. This analysis was
performed in \cite{mo1:pre96} in the context of reaction-diffusion
equations of activator-inhibitor type. Here we will re-derive these
results using the interfacial approach [Eq.~(\ref{lin})].

\subsubsection{Spot}

\begin{figure}
  \centerline{\psfig{figure=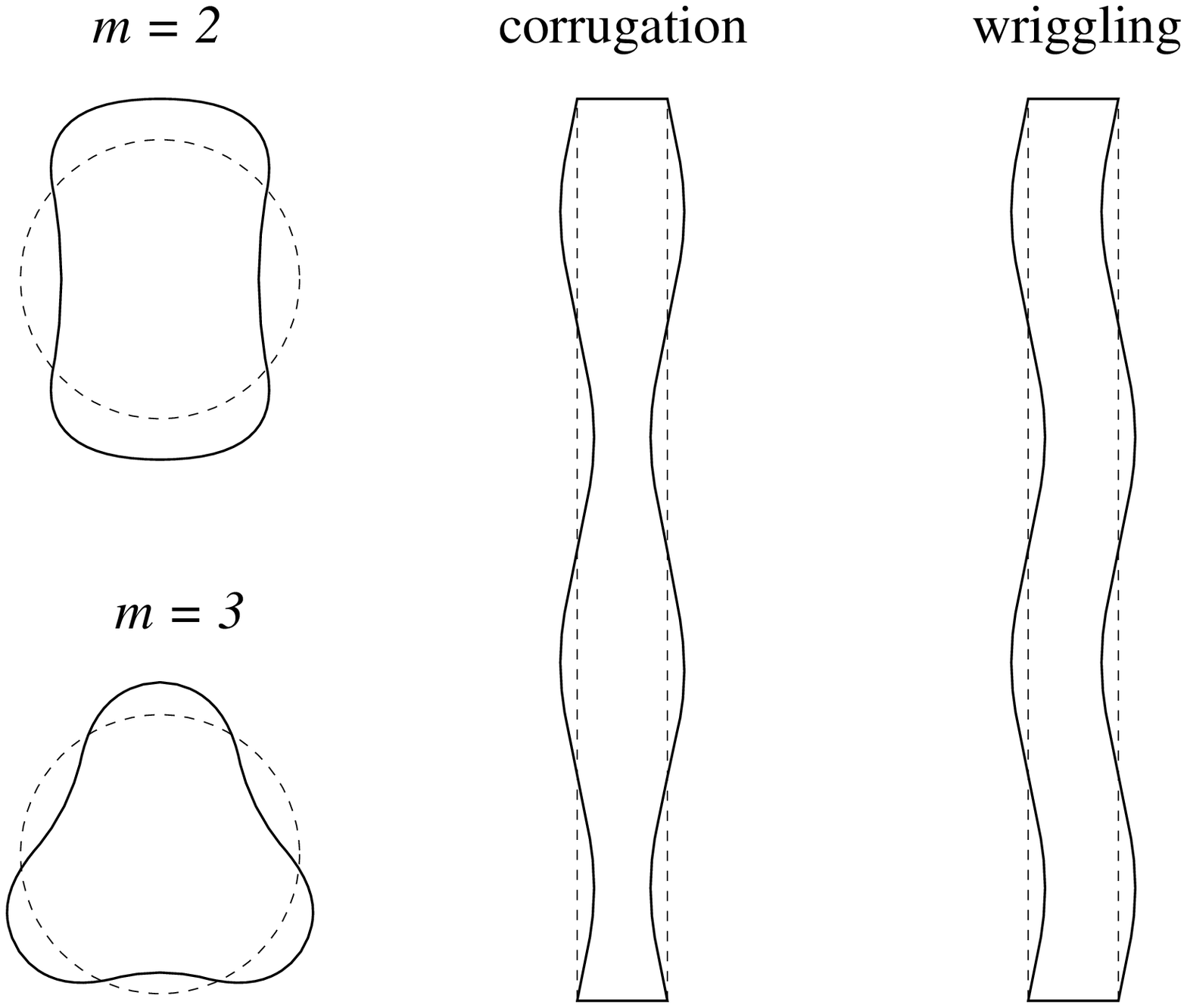,width=3in}}
 \caption{Morphological instabilities of spots and stripes.}
 \label{f:instabsol}
\end{figure}

Let us consider a single localized spot first. The fluctuations of the
spot's shape are the azimuthal distortions of its walls characterized
by the azimuthal number $m$ (Fig. \ref{f:instabsol}).
Because of the radial symmetry, the operator $L$ in Eq. (\ref{lin}) is
diagonal in the basis formed by the functions $e^{i m \varphi}$,
where $\varphi$ is the polar angle that represents a point on the
interface. For a spot of radius $\rs_s \ll \ep^{-1}$ one can neglect
screening and use $G$ in the integral for $m > 0$, since these
fluctuations do not change the overall area of the spot. Assuming
$\rho = e^{i m \varphi}$ and calculating the respective integrals,
we obtain
\begin{eqnarray}
 \label{disp:spot} \lambda = \lambda_0 + {\sigma_0 m^2 \over \rs_s^2}
 + {2 \ep^2 \rs_s \over m},
\end{eqnarray}
where $\lambda_0$ is a constant that comes from the curvature and the
$2 (\hat n \cdot \vec\nabla \psi)$ term. In fact, we do not need to
calculate this constant from the definition. Instead, we can use the
translational symmetry of the problem and note that $\lambda = 0$ for
$m = 1$ to find that
\begin{eqnarray}
\lambda_0 = - {\sigma_0 \over \rs_s^2} - 2 \ep^2 \rs_s.
\end{eqnarray}
Note that $\lambda_0 < 0$ and is responsible for the instability of
the spot for large enough $\rs_s$ (see also \cite{ko:book,ohta89}).

According to Eq.~(\ref{disp:spot}), a single localized spot becomes
unstable ($\lambda < 0$) with respect to the $m$-th mode when $\rs_s >
\rs_{cm}$, where 
\begin{eqnarray}
 \label{rcm2} \rs_{cm} = \left( {\sigma_0 m (m + 1) \over 2 \ep^2}
 \right)^{1/3}.
\end{eqnarray}
The instability is realized first with respect to the fluctuation with
$m = 2$, so the spot is always unstable when $\rs_s > \rs_{c2}$, where
\begin{eqnarray}
 \label{rc22} \rs_{c2} = (3 \sigma_0)^{1/3} \ep^{-2/3}.
\end{eqnarray}
The $m = 0$ case can be treated analogously, this leads once again to
Eqs.~(\ref{Rmin}) and (\ref{deltab}).  Therefore, comparing
Eq.~(\ref{rc22}) with Eq.~(\ref{rsd}), we see that the spot can be
stable only when
\begin{eqnarray} \label{dbc2}
\delta_b < \delta < \delta_{c2}, ~~~\delta_{c2} = 3^{-1/3} \kappa^2
\sigma_0^{2/3} \ep^{2/3} \ln \ep^{-1},
\end{eqnarray}
so the spots are stable only in the limited range of $\delta \sim
\ep^{2/3} \ll 1$ (apart from the logarithmic terms).

A similar analysis shows that a thin annulus of radius $\rs_a \ll
\ep^{-1}$ considered in Sec. \ref{s:sol} is always unstable with
respect to the $m = 2$ wriggling mode, so we do not present this
analysis in detail here.

\subsubsection{Stripe}

Let us now turn to the solitary stripe. Let us choose the reference
frame in such a way that the stripe is oriented along the $y$-axis in
the $zy$-plane. Because of the mirror symmetry of the stripe in the
$z$-direction, there are two basic types of fluctuations: the
symmetric and the antisymmetric distortions of the stripe walls, both
characterized by the transversal wave vector $k_\perp$ (Fig.
\ref{f:instabsol}). Because of the translational symmetry in the
$y$-direction the operator in Eq. (\ref{lin}) is diagonal in
$k_\perp$. Assuming that $\rho^+ = e^{i k_\perp y}$ and $\rho^- = \pm
\rho^+$, where $\rho^\pm$ are the positions of the right and left
boundaries of the stripe, respectively, we can calculate the integral
in Eq.~(\ref{lin}) at the location of the right wall:
\begin{eqnarray}
 \label{Rpmk} 4 \ep^2 \oint dS' G_\ep(x - x') \rho(x') && \nonumber \\
 && \hspace{-4cm} = {2 \ep^2 [1 \pm \exp( - \ls_s \sqrt{\ep^2 \kappa^2
 + k_\perp^2} ) ] \over \sqrt{\ep^2 \kappa^2 + k_\perp^2} } e^{i
 k_\perp y},
\end{eqnarray}
where ``+'' corresponds to the symmetric, and ``--'' to the
antisymmetric fluctuations, respectively; $\ls_s$ is the width of the
stripe.

For the stripe, the curvature terms in Eq.~(\ref{lin}) are zero, and
the $2 (\hat n \cdot \vec\nabla \psi)$ term reduces to a constant
$\lambda_0 < 0$. The case of the symmetric and antisymmetric
fluctuations must be treated separately. For $k_\perp \gg \ep$ the
expression for $\lambda = \lambda_+$ for the symmetric fluctuation is
(to the leading order in $\ep$)
\begin{eqnarray}
 \label{stripedis} \lambda_+ = \lambda_0 + \sigma_0 k_\perp^2 + {2
 \ep^2 [ 1 + \exp ( - k_\perp \ls_s)] \over k_\perp} .
\end{eqnarray}
On the other hand, when $\ls_s \ll \ep^{-1}$ and $k_\perp \ls_s \ll
1$, to the leading order in $\ep$ the expression for $\lambda =
\lambda_-$ for the antisymmetric fluctuation becomes
\begin{eqnarray}
 \label{anti} \lambda_- = \lambda_0 + \sigma_0 k_\perp^2 + 2 \ep^2
 \ls_s - \ep^2 \ls_s^2 \sqrt{\ep^2 \kappa^2 + k_\perp^2}.
\end{eqnarray}
Once again, we can use translational symmetry in the $z$-direction to
calculate $\lambda_0$, since $\lambda_- = 0$ when $k_\perp = 0$. We
get
\begin{eqnarray} \label{lam0str}
\lambda_0 = - 2 \ep^2 \ls_s + \ep^3 \kappa \ls_s^2 + O(\ep^4
\ls^3).
\end{eqnarray}

The analysis of the transcendent Eq.~(\ref{stripedis}) with
$\lambda_0$ given by the first term in Eq.~(\ref{lam0str}) shows that
the instability of the stripe with respect to the symmetric
distortions of its walls (corrugation) with $k_\perp = k_c$ occurs at
$\ls_s > \ls_{c1}$, where \cite{mo1:pre96}
\begin{eqnarray}
 \label{corr} k_c = 1.13 \sigma_0^{-1/3} \ep^{2/3}, ~~~\ls_{c1} = 1.66
 \sigma_0^{1/3} \ep^{-2/3}.
\end{eqnarray}

According to Eq.~(\ref{anti}), the stripe becomes unstable with
respect to the antisymmetric distortions of its walls (wriggling) at
$k \rightarrow 0$ and $\ls_s > \ls_{c2}$, where
\begin{eqnarray}
 \label{wrig} \ls_{c2} = (2 \sigma_0 \kappa)^{1/2} \ep^{-1/2}.
\end{eqnarray}
This is also clear from Eq.~(\ref{F:stripe}). Comparing
Eq. (\ref{wrig}) and (\ref{corr}), one can see that the instability
with respect to wriggling is realized before the instability with
respect to the corrugation. In view of Eq.~(\ref{ls}), the stripe is
stable only when
\begin{eqnarray}
\ep \ln \ep^{-1} \lesssim \delta < \delta_\perp,
\end{eqnarray}
where $\delta_\perp$ is defined in Eq.~(\ref{deltaperp}).  Thus, the
region of existence of stable stripes is wider than that for spots,
see Eq.~(\ref{dbc2}).

\subsection{Morphological instabilities of hexagonal and lamellar
patterns} \label{s:perstab}

The solution of Eq. (\ref{lin}) in the case of an arbitrary
multidomain pattern is a formidable task. However, a simplification of
this problem is possible in the case of periodic patterns. Then, by
Bloch theorem, the operator $L$ can be partially diagonalized by
considering the fluctuations modulated by the wave vector $\mathbf{k}$
which lies in the first Brillouin zone of the underlying domain
lattice. The situation here is not unlike the problem of finding the
band structure of a crystal \cite{ziman}. Below, we consider stability
of the hexagonal and lamellar patterns in two dimensions.

\subsubsection{Hexagonal pattern}

Let us consider a hexagonal pattern of period $\ls_p$ made of circular
domains of radius $\rs_s$. For each domain centered at $\mathbf{R}_n$
let us write the displacement $\rho_n$ as
\begin{eqnarray}
\rho_n(\varphi) = \sum_m a_m e^{i \mathbf{k} \cdot \mathbf{R}_n - i m
\varphi},
\end{eqnarray}
where the angle $\varphi$ represents a point on the interface of each
domain. Equation (\ref{lin}) for the fluctuations with a given
$\mathbf{k}$ in the first Brillouin zone then reduces to
\begin{eqnarray}
 \label{disp:hex} \left( {\sigma_0 m^2 \over \rs_s^2} + \lambda_0
-\lambda \right) a_m = - \sum_{m'} {\sf R}_{mm'} (\mathbf k) a_{m'},
\end{eqnarray}
where $m$ and $m'$ are the azimuthal numbers, $\lambda_0$ is a
constant independent of $m$ and $\mathbf k$ (assuming that with good
accuracy $\hat n \cdot \nabla \psi$ is radially-symmetric in the
interface), and ${\sf R}_{mm'}(\mathbf{k})$ are the
$\mathbf{k}$-dependent matrix elements of $4 \ep^2 G_\ep(x - x')$. A
calculation of Appendix \ref{a:hexstab} shows that \cite{m1:prl97}
\begin{eqnarray}
 \label{Rmmk} {\sf R}_{mm'}(\mathbf k) = \frac{16 \pi \ep^2
 \rs_s}{\ls_p^2 \sqrt{3}} \sum_n {e^{i (m - m')
 \left(\vartheta_{\mathbf{k} + \mathbf{k}_n} + \frac{\pi}{2}\right)}
 \over |\mathbf{k} + \mathbf{k}_n|^2 + \ep^2 \kappa^2} && \nonumber \\
 && \hspace{-6cm} \times J_m (|\mathbf{k} + \mathbf{k}_n| \rs_s)
 J_{m'} (|\mathbf{k} + \mathbf{k}_n| \rs_s),
\end{eqnarray}
where $\mathbf{k}_n$ run over the reciprocal lattice, $J_m(x)$ are the
Bessel functions of the first kind and $\vartheta_{\mathbf{k} +
\mathbf{k}_n}$ is the angle between the vector $\mathbf{k} +
\mathbf{k}_n$ and the $x$-axis.

The value of $\lambda_0$ can be calculated by noting that the
translational invariance of the system requires that $\lambda = 0$ for
$\mathbf k = \mathbf 0$ and $m = 1$, so (Appendix \ref{a:hexstab}) 
\begin{eqnarray}
 \label{l0:hex} \lambda_0 = - {\sigma_0 \over \rs_s^2} - 2 \ep^2 \rs_s
 \left( 1 - { 2 \pi \rs_s^2 \over \sqrt{3} \ls_p^2} \right).
\end{eqnarray}
In writing the above equations we assumed that $\rs_s \sim \ls_p \ll
\ep^{-1}$.

\begin{figure}
 \centerline{\psfig{figure=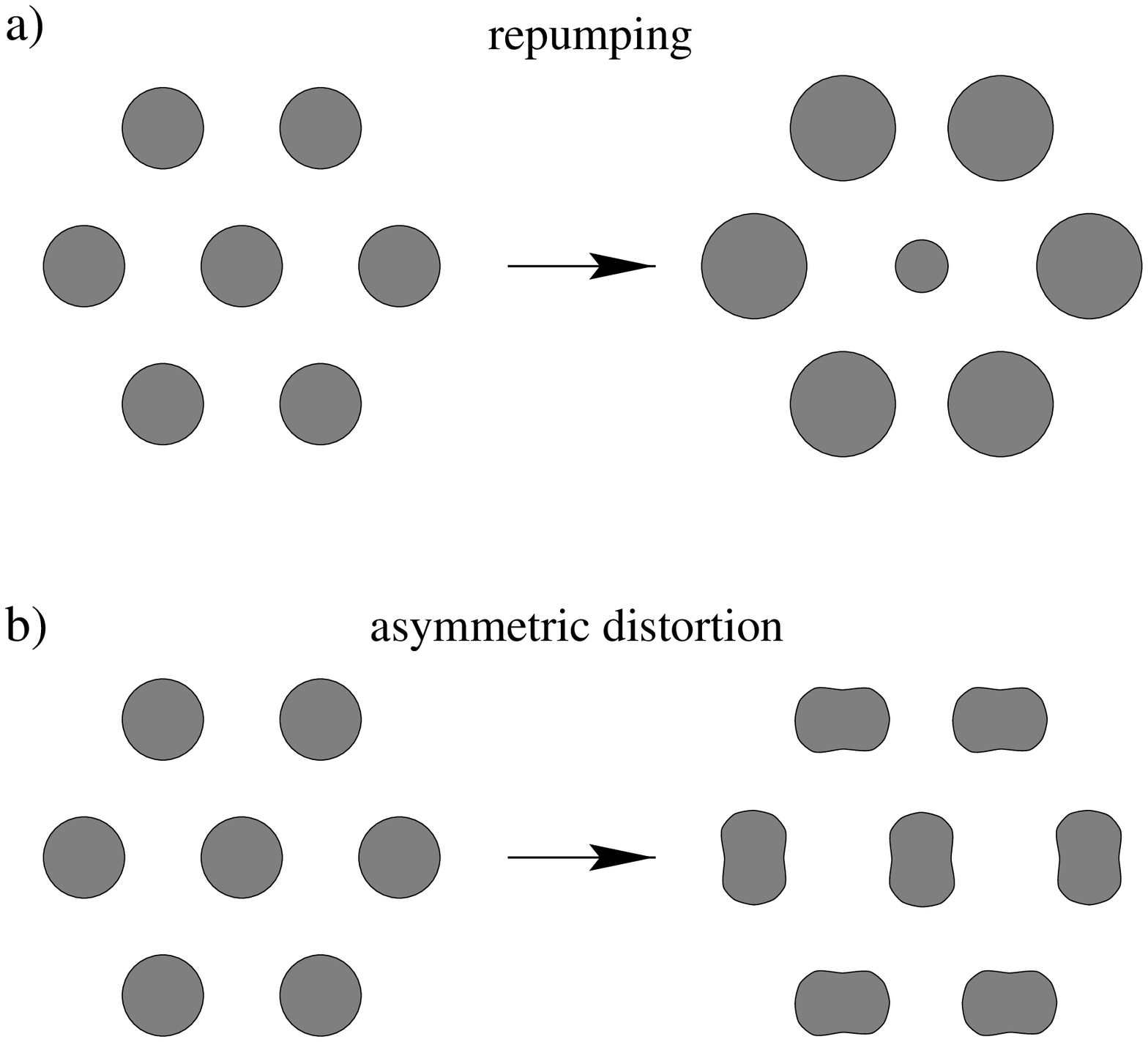,width=3in,angle=-90}}
 \caption{Two major types of instabilities of the hexagonal pattern.}
 \label{f:instabhex}
\end{figure}

\begin{figure}[b]
\centerline{\psfig{figure=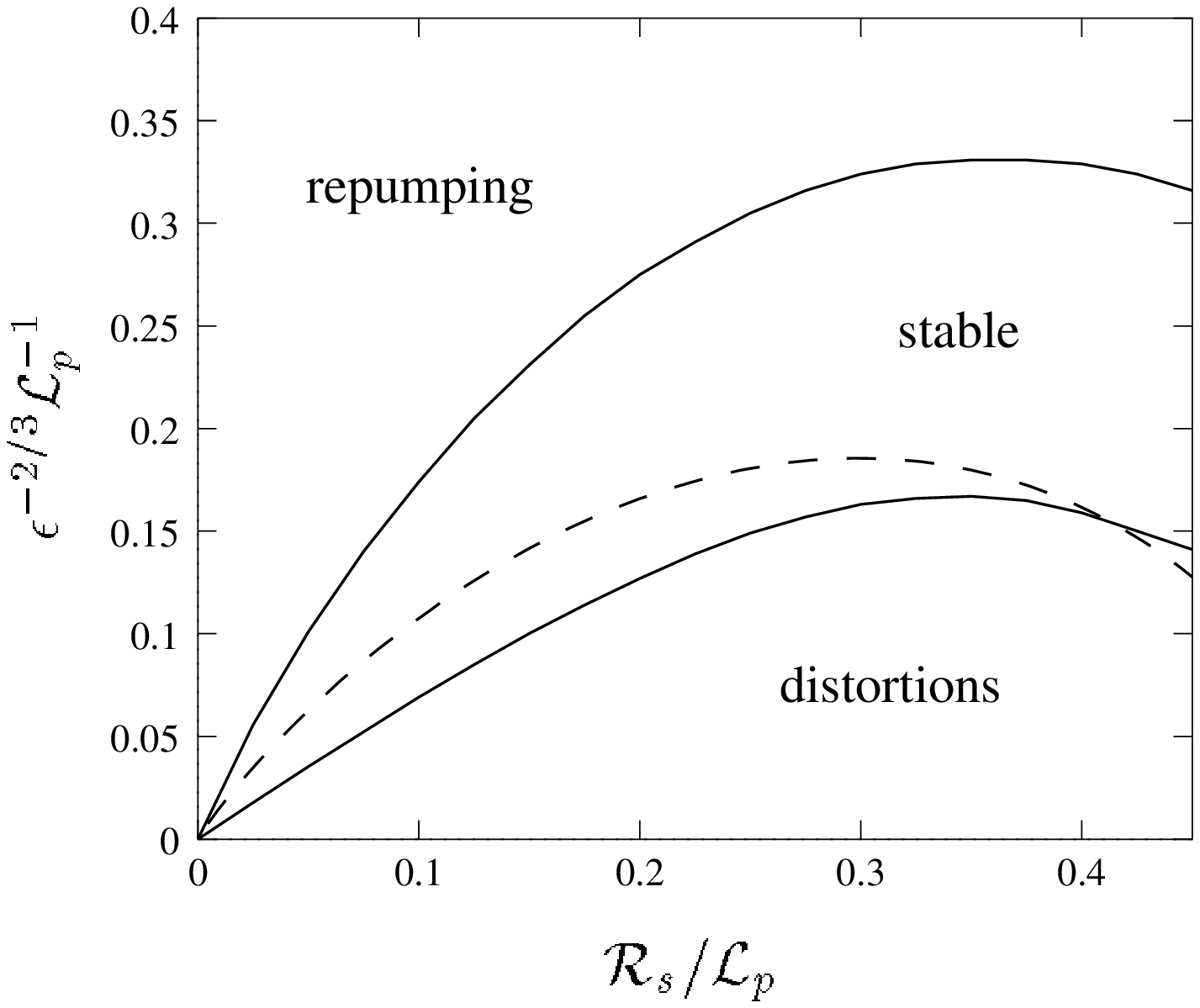,width=3in}}
\caption{The stability diagram for the hexagonal pattern with
$\sigma_0 = 2 \sqrt{2} / 3$. The dashed line corresponds to the
hexagonal pattern with the lowest free energy given by
Eq.~(\ref{lp2eq}).}  \label{f:phasehex}
\end{figure}

As was shown qualitatively by Kerner and Osipov, for the most
dangerous fluctuations the wave vector $\mathbf{k}$ will lie close to
the edge of the Brillouin zone \cite{ko:book}. There are two basic
types of fluctuations we need to consider: the fluctuations which lead
to repumping of the order parameter between the neighboring domains
(Fig. \ref{f:instabhex}[a])
and the fluctuations which lead to the asymmetric distortions of the
domains (Fig. \ref{f:instabhex}[b]) ( \cite{ko:book}, see also
Sec. \ref{s:fluct}).  The analysis of Eq. (\ref{disp:hex}) shows that
the most dangerous fluctuations leading to repumping have $\mathbf{k}
= \frac{1}{3} (\mathbf{b}_1 - \mathbf{b}_2)$, where $\mathbf{b}_1$ and
$\mathbf{b}_2$ are the reciprocal lattice vectors which make a
$120^{\rm o}$ angle (see Appendix \ref{a:hexstab}), while the most
dangerous fluctuations leading to distortions have $\mathbf k =
\frac{1}{2} (\mathbf b_1 + \mathbf b_2)$ \cite{m1:prl97}. The
instability $\lambda < 0$ occurs with respect to repumping when $\ls_p
< {\cal L}_{p0}$ or with respect to the asymmetric distortion when
$\ls_p > {\cal L}_{p2}$, where ${\cal L}_{p0,2}$ depend on $\ep$ and
$\rs_s / \ls_p$. The resulting stability diagram obtained by numerical
solution of Eq.  (\ref{disp:hex}) with the parameters of the
mean-field model from Sec. \ref{s:mf} is presented in Fig.
\ref{f:phasehex}.  From this figure one can see that only the patterns
with the period $\ls_{p0} < \ls_p < \ls_{p2}$ are stable. 

Figure \ref{f:phasehex} also shows the period of the lowest free
energy hexagonal pattern given by Eq. (\ref{lp2eq}). One can see that
this pattern is stable for all values of $\rs_s / \ls_p$ (except,
possibly, for $\rs_s / \ls_p$ close to 0.5 where the assumption about
the circular shape of the domains ceases to be valid).  As was noted
in Sec. \ref{s:hexlam}, in two dimensions the lowest free energy
hexagonal pattern is expected to be the global minimizer of the free
energy if $f < 0.35$, or, equivalently, if $\rs_s / \ls_p < 0.31$,
whereas for $0.31 < \rs_s / \ls_p < 0.37$ (the second condition means
that $\phb < 0$) the global minimizer should be the lamellar
pattern. Figure \ref{f:phasehex}, however, does not show the
transition from the hexagonal to the lamellar pattern, so in fact the
lowest energy hexagonal pattern is at least {\em metastable} for all
values of $\phb$ at which it exists.

\subsubsection{Lamellar pattern}

In the case of the lamellar pattern one should consider the
fluctuations that are modulated by the wave vector $k_\parallel$ from
the first Brillouin zone in the direction along the normal to the
stripes and an arbitrary wave vector $k_\perp$ in the transverse
direction:
\begin{eqnarray} \label{fluct:lam}
\rho_n^\pm = \rho_0^\pm e^{i k_\perp y + i k_\parallel n \ls_p},
\end{eqnarray}
where $\rho_n^\pm$ are the displacements of the right and left walls
of the stripe in the $n$-th period of the lamellar pattern at $y =
0$. Because of the translational symmetry in the direction along the
stripes, these fluctuations are the eigenfunction of $L$ in
Eq. (\ref{lin}). One can then reduce the operator $L$ to a $2 \times
2$ matrix, so after a tedious calculation (Appendix \ref{a:lamstab},
see also \cite{ko:book,pismen94})
\begin{eqnarray}
 \label{disp:lam} \lambda_\pm = \lambda_0 + \sigma_0 k_\perp^2 +
 \mathsf R_\pm(k_\parallel, k_\perp),
\end{eqnarray}
where $\mathsf R_\pm(k_\parallel, k_\perp)$ are given by
\begin{eqnarray}
\label{Rpmlam}
\mathsf R_\pm(k_\parallel, k_\perp) = {4 \ep^2 e^{k \ls_p} \over k
\left(1 - 2 e^{k \ls_p} \cos k_\parallel \ls_p + e^{2 k \ls_p} \right)
} \times \nonumber \\ \left\{ \sinh k \ls_p \pm \bigl[ \left( \sinh[
k(\ls_p - \ls_s)] + \cos k_\parallel \ls_p \sinh k \ls_s \right)^2
\right. \nonumber \\ \left. + \sin^2 k_\parallel \ls_p \sinh^2 k \ls_s
\bigr]^{1/2} \right\}, \nonumber \\
\end{eqnarray}
where $k = \sqrt{\ep^2 \kappa^2 + k_\perp^2}$. As before, the value of
$\lambda_0$ is determined with the aid of the translational invariance
of the system, which requires that $\lambda = 0$ for $k_\parallel =
k_\perp = 0$ for the antisymmetric fluctuation.  This gives the
following value of $\lambda_0$ (to the leading order in $\ep$):
\begin{eqnarray}
 \label{l0:lam} \lambda_0 = - 2 \ep^2 \ls_s \left(1 -
 \frac{\ls_s}{\ls_p} \right).
\end{eqnarray}

The analysis of Eq. (\ref{disp:lam}) shows that (in the validity range
of Eq.~(\ref{Rpmlam}), that is, when $\ls_p \gg \ln \ep^{-1}$, see
also \cite{ko:book}) the repumping instability is not realized for the
lamellar pattern. This can be explained by a simple argument: the
curvature of the stripes is equal to zero, so there is no force that
would lead to the domain collapse as in the case of the spot. The
analysis of Eq. (\ref{disp:lam}) also shows that the most dangerous
fluctuation leads to the antisymmetric distortions of the stripe and
has $k_\parallel = 0$ and $k_\perp \rightarrow 0$ (see also
\cite{pismen94,ren02}). All other instabilities, such as the
corrugation instability, occur at higher values of $\ls_p \sim
\ep^{-2/3}$ (compare with \cite{qi97,yeung96}). Also notice that when
$k_\perp = 0$, what corresponds to the one-dimensional situation, the
lamellar pattern is always stable when $\ln \ep^{-1} \ll \ls_p \ll
\ep^{-1}$ (see Appendix \ref{a:lamstab}). This is in agreement with
the result of Ren and Wei that in this situation the lamellar patterns
are all local minimizers of Eq.~(\ref{F:inter}) \cite{ren00}.

\begin{figure}[b]
  \centerline{\psfig{figure=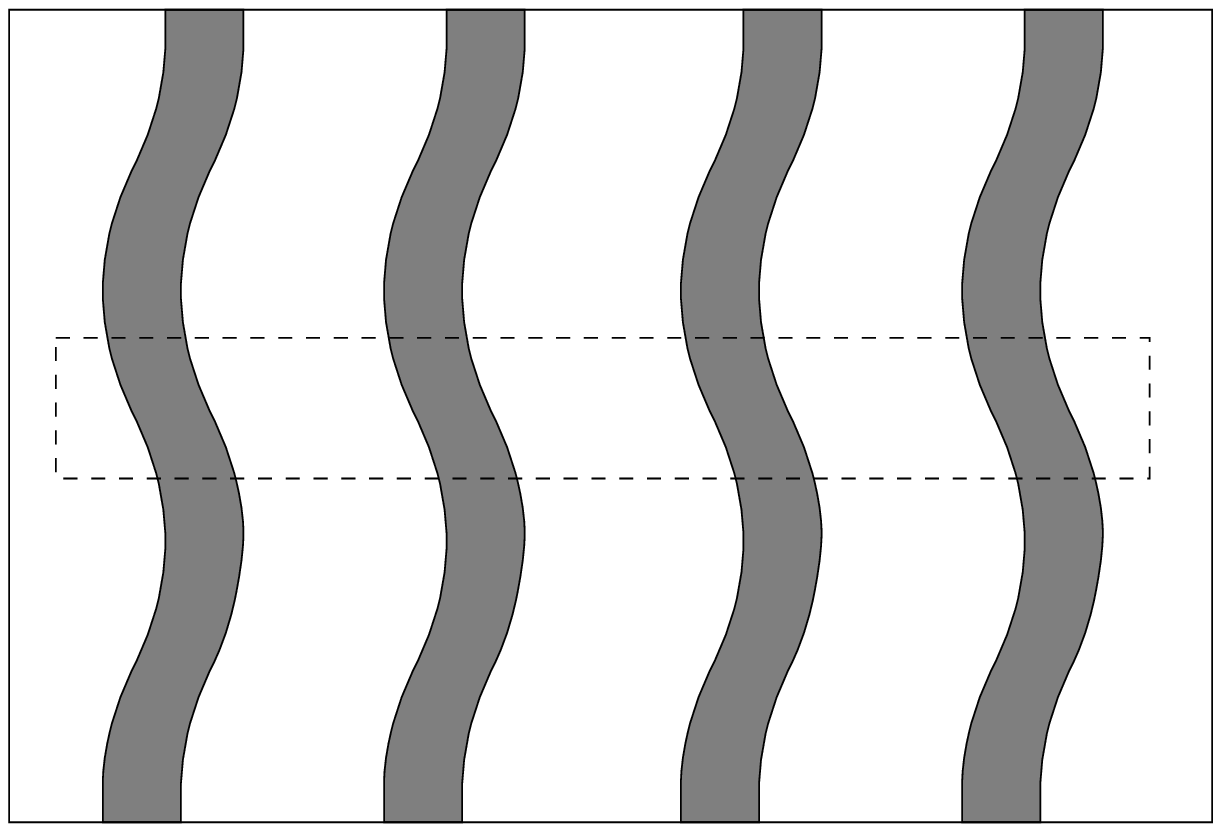,width=3in}}
 \caption{Wriggled lamellar pattern.}
 \label{f:wriglam}
\end{figure}

Solving Eq. (\ref{disp:lam}) with $k_\parallel = 0$, we obtain that
the instability is realized when $\ls_p = \ls_p^*$, where $\ls_p^*$ is
the period of the lowest free energy lamellar pattern given by
Eq. (\ref{lp1eq}) (see Appendix \ref{a:lamstab}). This result was also
obtained by Yeung and Desai in the case $f = 0.5$
\cite{yeung94}. Thus, the lowest free energy lamellar pattern is {\em
marginally} stable with respect to the wriggling instability. This
fact has a simple geometric interpretation, and should in fact be true
for any system with long-range competing interactions. Indeed,
consider a small wriggling modulation of the lamellar pattern
(Fig. \ref{f:wriglam}).
Inside the dashed line rectangle the stripes can be considered as
straight, but rotated by a small angle $\vartheta$. This pattern can
be again considered as a lamella, but with a smaller period $\ls_p' =
\ls_p \cos \vartheta$. The free energy of the system in this case will
increase if $\ls_p < \ls_p^*$, since the free energy is a decreasing
function of $\ls_p$ for these values of $\ls_p$, or decrease if $\ls_p
> \ls_p^*$, since there the free energy is an increasing function of
$\ls_p$. The case $\ls_p = \ls_p^*$ is marginal. Therefore, the
lamellar pattern will be unstable with respect to wriggling if $\ls_p
> \ls_p^*$ or stable otherwise. 

It is interesting to note the analogy between the lamellar patterns
and smectic A phases. In smectics the long-wave modulations of the
layered structure cost free energy $\Delta F \propto \left( \bar B
k_\parallel^2 + K_1 k_\perp^4 \right) |u_z(k_\parallel, k_\perp)|^2$,
where $u_z(k_\parallel, k_\perp)$ is the amplitude of the layer
displacements modulated by wave vecotors $k_\parallel$ and $k_\perp$
along and perpendicular to the layers, respectively
\cite{prost,landau5}. This is precisely what we get for the pattern
with $\ls_p = \ls_p^*$ in the limit of small $k_\parallel$, $k_\perp$
(see Appendix \ref{a:lamstab}). Furthermore, in view of this analogy
the long-wave instability of the lamellar patterns with $\ls_p >
\ls_p^*$ is equivalent Helfrich-Hurault instability of smectics under
stretching deformations. Also note that under the influence of thermal
fluctuations the lamellar pattern with $\ls_p = \ls_p^*$ is subject to
the Landau-Peierls instability \cite{prost,landau5}. We would like to
point out, however, that all this does not apply to the metastable
lamellar patterns with $\ls_p < \ls_p^*$, which, according to our
calculations, have finite shear modulus.

\section{Scenarios of domain pattern formation}

The analysis in the preceding sections shows that the patterns in
systems with long-range Coulomb interactions are very sensitive to the
parameters $\phb$ and $\ep$ and 
\begin{figure}
 \centerline{\psfig{figure=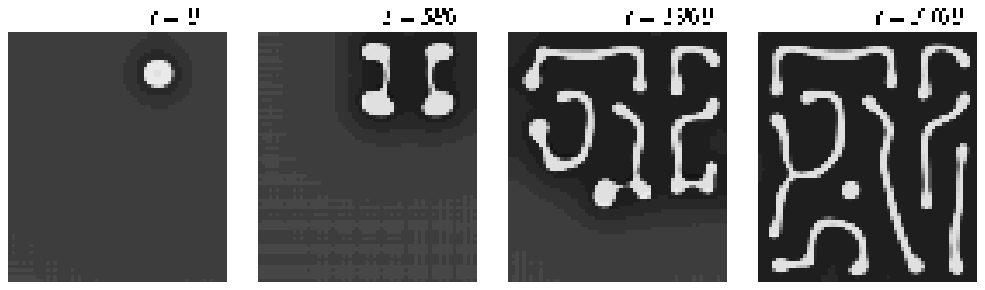,width=5.5in}}
 \caption{Formation of a complex pattern as a result of an instability
 of a spot. Results of the numerical solution of Eqs.~(\ref{dyn}) with
 $\ep = 0.025$ and $\phb = -0.6$, with no-flux boundary
 conditions. The system is $400 \times 460$. } \label{f:split}
\end{figure}
can undergo various instabilities. As the temperature is lowered, both
$\ep$ and $|\phb|$ rapidly decrease, see, for example,
Eqs.~(\ref{scale:gen}) and (\ref{ep}). In this situation a small
variation of the temperature may trigger complex spatiotemporal
behavior in the system. Now we would like to ask the following
questions: how do the patterns form in the initially homogeneous
system and how do the patterns already present in the system react to
changes in the external parameters?  In principle, to answer these
questions we need to specify the dynamics of patterns. This question
is quite complicated and significantly depends on a particular system,
despite the universality possessed by the free energy (for various
examples, see \cite{oono88,yeung94,m:phd,ko:book}). However, if the
dynamics of the system is {\em dissipative}, it will result in the
decrease of the free energy of the patterns with time. Note that since
the free energy functionals in Eqs.~(\ref{F}) or (\ref{F:inter}) are
bounded from below in systems of finite volume, in the absence of the
noise the patterns must evolve to local minimizers of the free
energy. To mimic this behavior, we will use the simple gradient
descent dynamics defined by
\begin{eqnarray} \label{dyn}
{\partial \phi \over \partial t} = - {\delta F \over \delta \phi},
\end{eqnarray}
where $F$ is given by Eq.~(\ref{F}). This evolution equation is in
fact applicable to a number of systems with non-conserved order
parameter \cite{grousson01,mamin94,ko:book,cross93,kapral}. However,
our conclusions should not qualitatively depend on this particular
choice, since the evolution of patterns will generally be guided by
the free energy landscape and the morphological instabilities of the
patterns. Note that Eq.~(\ref{dyn}) is equivalent to a
reaction-diffusion system with the fast inhibitor and can be reduced
to a free boundary problem in the limit $\ep \rightarrow 0$
\cite{m:pre96}, which, in turn, is the gradient descent dynamics for
the interfacial free energy \cite{goldstein96}.

\subsection{Nucleation} \label{s:nucl}

The first question is how the domain patterns form in the system in
the first place. As was discussed in Sec. \ref{s:mf}, at sufficiently
high temperatures above $T_c$ the homogeneous phase is the only
equilibrium state. In the mean-field model of Sec. \ref{s:mf} the
homogeneous phase becomes unstable as the temperature of the system is
lowered. At $\phb = 0$ this happens when $\ep(\tau) = \ep_c \sim
1$. On the other hand, when the original (unscaled) value of $\phb$ is
different from 0, the homogeneous state will remain stable even for
lower temperatures. The greater the (unscaled) value of $|\phb|$, the
lower the temperature at which the homogeneous phase will lose its
stability. This means that for $\al \ll 1$ the homogeneous phase will
typically remain stable in a range of $\tau$ for which $\ep \ll 1$.

On the other hand, as was shown in Sec. \ref{s:sol}, when the scaled
value of $|\phb|$ is less than $|\phb_b| = 1 - \delta_b$, in addition
to the stable homogeneous phase the system can support stable domain
patterns (spots). In a narrow range of $|\phb_m| < |\phb| < |\phb_b|$,
where $|\phb_m| = 1 - \delta_m$, the spots will be energetically
unfavorable. On the other hand, in a wide range of $|\phb| < |\phb_m|$
the domain patterns will have lower free energy than the homogeneous
phase. In the mean-field model of Sec. \ref{s:mf} the homogeneous
phase remains stable as long as $|\phb| > |\phb_c| \simeq 1/\sqrt{3}$
for $\ep \ll 1$. Therefore, at $|\phb_c| < |\phb| < |\phb_m|$ the
homogeneous phase is {\em metastable}.

The metastability of the homogeneous phase implies the possibility of
{\em nucleation} of the domain patterns as a result of thermal
fluctuations. It is natural to assume that the nucleating droplet in
two dimensions is a spot (it is localized in space and
radially-symmetric). Let us consider the nucleation of the positive
domains from the negative homogeneous phase. In this case the value of
$\phb$ is negative, so, as the temperature of the systems decreases,
the value of $\phb$, as well as $\delta$ [recall Eq.~(\ref{delta})],
increases. As was shown in Sec. \ref{s:sol}, at $\delta > \delta_b$
there are two spot solutions. The spot with the radius $\rs = \rs_n$
is in the unstable equilibrium with the homogeneous phase (see
Fig. \ref{f:rspot}). Therefore, it is this solution that should play
the role of the nucleation droplet in our system.  Note that the
radius of the nucleation droplet $\rs_n < \rs_\mathrm{min}$ and is
bounded for all $\delta$, in contrast to systems with first-order
phase transitions. This is a distinctive property of the systems with
long-range Coulomb interactions.

According to Eq.~(\ref{Rspot}), for $\delta$ close to $\delta_b$ the
free energy cost of the nucleation droplet is (apart from a weak
logarithmic dependence) 
\begin{eqnarray}
 \label{dropletcost} \Delta F_\mathrm{drop} \propto \ep^{-2/3} \gg 1.
\end{eqnarray}
Note that in three dimensions the same arguments give $\Delta
F_\mathrm{drop} \propto \ep^{-4/3} \gg 1$ also. Since this free energy
barrier is high, the arguments of the nucleation theory apply here.

In the narrow range of $\delta_b < \delta < \delta_m$, a nucleation
event will result in the formation of a stable spot whose free energy
is higher than that of the homogeneous phase. Therefore, in this
situation the spot itself will be metastable and will decay back into
the homogeneous phase. However, when $\delta$ is not in the immediate
vicinity of $\delta_b$, the free energy barrier the spot has to
overcome to decay will also scale as in Eq.~(\ref{dropletcost}), so
such spots will be long-lived metastable states that can be excited by
thermal fluctuations. Therefore, the thermodynamic equilibrium state
of the system for these parameters is a rarefied gas of noninteracting
spots. In this situation the spots will play the role of {\em
quasiparticles}.

When $\delta$ exceeds $\delta_m$, the spot becomes thermodynamically
more favorable than the homogeneous phase. For $\delta_m < \delta <
\delta_{c2}$ the spots with the radius $\rs_s$ are stable, so a
nucleation event will result in the formation of a single stable spot.
This is another distinctive feature of nucleation in our system:
sufficiently close to $\delta_b$ a single nucleation event will result
in a formation of only one spot. However, in order for the system to
come to the equilibrium, it has to become filled with spots, so the
transition from the metastable homogeneous phase to the equilibrium
multidomain pattern requires many nucleation events. These events will
occur on an extremely long time scale $\tau_\mathrm{nucl} \sim e^{c
\Delta F_\mathrm{drop}} \gg \tau_\mathrm{rel}$, where $c \gtrsim 1$ is
a constant, $\tau_\mathrm{rel}$ is a characteristic system's
relaxation time. Dynamically, this phenomenon can be identified as
{\em aging} \cite{bouchaud97}.

According to Eq. (\ref{rsd}), as the temperature decreases, and,
therefore, $\phb$ and $\delta$ increase, the radius of the nucleation
droplet gets smaller (in the scaled units), while the radius of the
stable spot becomes larger, so at $\delta = \delta_{c2}$ the spot with
the radius $\rs_s$ will become unstable with respect to the
morphological instability (see Sec. \ref{s:morsol}).  In this case the
nucleation scenario will change. Instead of a single spot, a more
complex pattern will form as result of a single nucleation event (see
Sec. \ref{s:growth}). At the same time, the nucleation barrier
decreases with temperature. At $\delta \sim 1$ the free energy barrier
becomes $\Delta F_\mathrm{drop} \sim 1$ [see Eqs. (\ref{Rspot}) and
(\ref{rsd})]. So, in the renormalized model of Sec. \ref{s:renorm}
nucleation becomes meaningless for these values of $\delta$, and one
can talk about the {\em instability} of the homogeneous phase. Let us
point out that for $\delta \gtrsim \delta_{c2}$ the annulus should
also be considered as a potential candidate for the nucleation
droplet. The comparison of the free energies of the annulus and the
spot of radius $\rs_n$ shows, however, that the spot always has lower
free energy.

Let us consider nucleation in the renormalized model of
Sec. \ref{s:renorm} in more detail. Depending on the cooling rate,
several situations are possible. When the cooling rate is very small
(with the characteristic time scale much longer than
$\tau_\mathrm{nucl}$), the system will have enough time to equilibrate
and will get filled with spots. If the cooling rate is such that its
characteristic time scale is comparable to $\tau_\mathrm{nucl}$, the
system will enter into the aging regime. Notice that these phenomena
will occur only in a narrow range of temperatures at which $\delta
\sim \delta_b$.

The situation will change qualitatively when $\delta > \delta_{c2}$,
when the temperature falls below the value at which the spot becomes
unstable with respect to a morphological instability. In this case a
single nucleation event will produce a spot which will further develop
into a more complex extended pattern, filling up the whole system (see
Sec. \ref{s:growth}). The time scale of this process is
$\tau_\mathrm{rel}$ and is much shorter compared to
$\tau_\mathrm{nucl}$, so, if the cooling rate is sufficiently fast,
only a single nucleation event is enough to create a pattern that will
fill the entire system. This will also be the case in the mean-field
model of Sec. \ref{s:mf}. Finally, if the cooling rate is very fast,
the system will not have enough time to nucleate even a single domain,
so it will enter the region in which the homogeneous state of the
system is unstable. In that case a pattern consisting of the domains
whose characteristic size is comparable with the correlation length
will form spontaneously and then evolve towards equilibrium via
coarsening (see Sec. \ref{s:coarse}).

If small local inhomogeneities exist in the system, they can work as
the nucleation centers. One could, for example, have a slightly
nonuniform distribution of $\phb$ across the system. If their
amplitude and size are not very large, the nucleation events will
produce stable spots that will be pinned to the locations of these
inhomogeneities. If, on the other hand, the amplitude and size of
these inhomogeneities are large enough, the spots that nucleate at
their locations may be unstable with respect to the morphological
instabilities, so they can work as the nucleation centers for the
spatially extended patterns.

\subsection{Growth of complex patterns} \label{s:growth}

As was discussed in Sec. \ref{s:nucl}, typically a spot that forms as
a result of a nucleation event will be unstable with respect to the
deformations of its shape. After such a spot is formed, it will start
to grow into a more complex pattern on the time scale of
$\tau_\mathrm{rel} \ll \tau_\mathrm{nucl}$. Therefore, in the process
of growth of such a spot thermal fluctuations become unimportant. A
typical evolution of a pattern in this situation is shown in
Fig. \ref{f:split} which shows the result of the numerical solution of
Eq.~(\ref{dyn}) in this parameter regime.  At long times, the system
evolves to a disordered metastable equilibrium pattern.

\begin{figure}
  \centerline{\psfig{figure=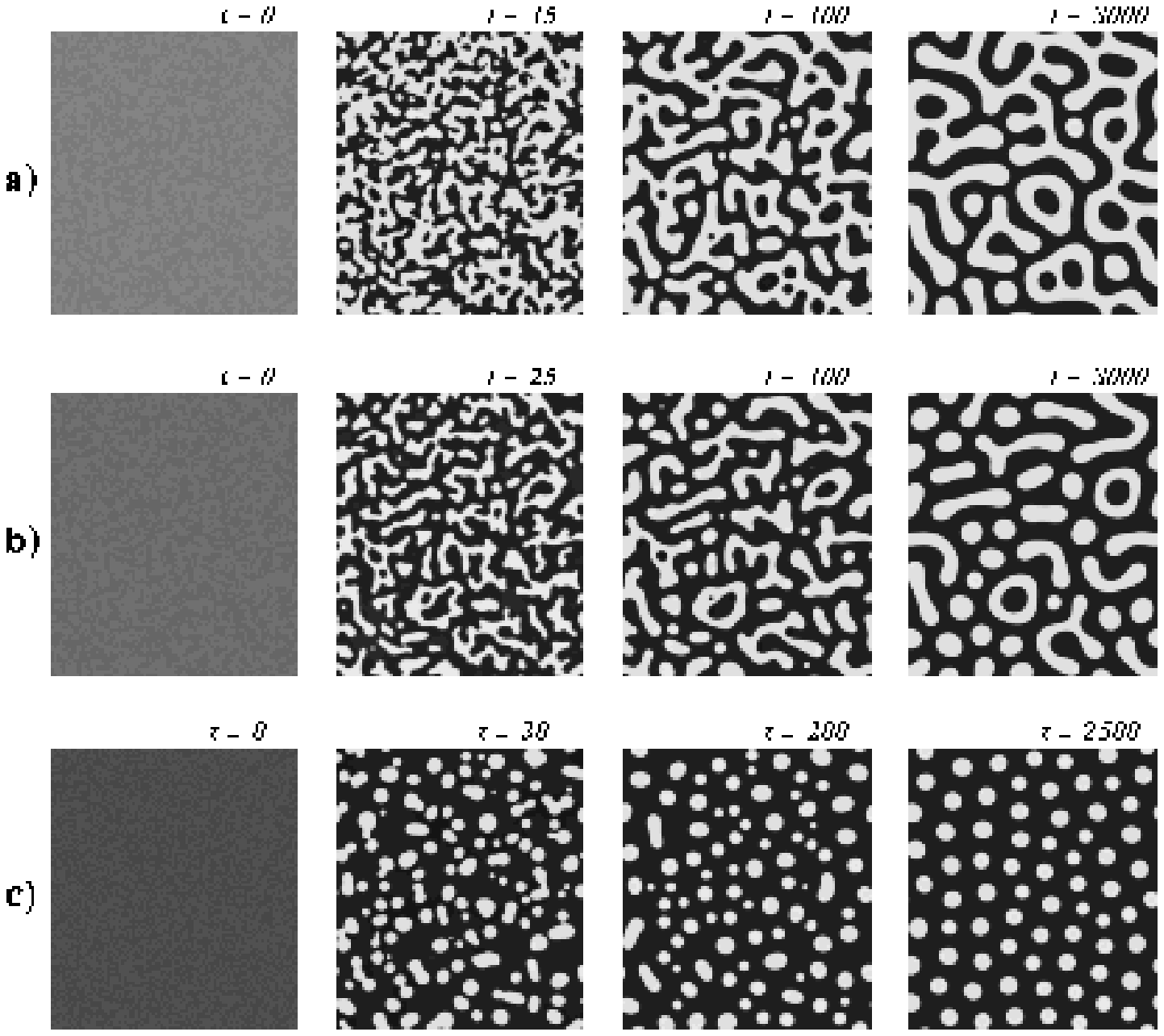,angle=-90,width=6in}}
  \caption{Coarsening of the domain patterns at different values of
  $\phb$: $\phb = 0$ (a), $\phb = -0.2$ (b), $\phb = -0.5$ (c).
  Results of the numerical simulations of Eqs.  (\ref{dyn}) with $\ep
  = 0.025$ and periodic boundary conditions. The system is $400 \times
  460$.}  \label{f:coarse1}
\end{figure}

The evolution of the unstable localized patterns within the framework
of Eq.~(\ref{dyn}) and the corresponding interfacial dynamics problem
was studied in detail in \cite{mo2:pre96,m:pre96,goldstein96}. The
solution of the interfacial dynamics equation shows that for
sufficiently large $\delta$ the morphological instability of a spot
will always lead to self-replication of spots \cite{m:pre96}. As a
result of the instability, the spot grows more and more distorted,
until at some point the interfaces touch, what leads to the pinch-off
and splitting (fission) of one domain into two. The daughter domains
move away from each other, and the process of splitting repeats
itself. This self-replication process will continue until the whole
system gets filled with the multidomain pattern
\cite{m:pre96,mo2:pre96}.

Note that these results hold in the asymptotic limit $\ep \rightarrow
0$, in which $\rs_s \sim \ep^{-2/3} \ll \ep^{-1}$, so the screening
effects may be ignored. On the other hand, for reasonably small but
finite $\ep$ the effective interaction may get truncated at distances
comparable to the sizes of the spots. In particular, as the spots move
away from each other after splitting, the interaction between the
distant portions of their interfaces may get screened, so the spots
will remain connected by a {\em stripe} as they move apart (see
Fig. \ref{f:split}, $t = 580$). For $\delta_{c2} < \delta \lesssim
\delta_{c3}$ (the latter corresponds to the value of $\delta$ at which
the spot becomes unstable with respect to the $m = 3$ mode) the tips
on both sides of the stripe will be stable, so as a result of the
destabilization a spot will transform into a stripe spanning across
the system. Note that according to Eq.~(\ref{F:stripe}), this can
happen only when $\delta > \delta_\perp$, when the stripe is
energetically favorable. According to Eqs.~(\ref{deltaperp}) and
(\ref{dbc2}), for not very small values of $\ep$ it may be possible to
have $\delta_{c2} \sim \delta_\perp$. Note that for these values of
$\delta$ the newly grown stripe will destabilize with respect to
wriggling and fill up the entire space of the system. Furthermore, the
stripe segments with the highest curvature may become unstable with
respect to fingering \cite{mo2:pre96}. This is what we see in the
numerical simulations of Eq.~(\ref{dyn}) in this parameter range. When
$\delta \gtrsim \delta_{c3}$, the tips of a stripe growing as a result
of a splitting event can further destabilize with respect to the $m =
3$ mode, what will result in tip splitting and the formation of a {\em
labyrinthine} pattern.  Note that similar results were obtained in the
case of reaction-diffusion systems with weak activator-inhibitor
coupling \cite{goldstein96}.

From the arguments above it is possible to conclude that following a
nucleation event at $\delta$ not very far from $\delta_{c2}$ and for
not very small values of $\ep$ the dominating pattern morphology is
that of the stripe.  Nevertheless, for larger values of $\delta$ the
spot morphology starts to compete with the stripe morphology. This is
because for large $\delta$ the interfaces will tend to split, so the
forming labyrinthine pattern will become {\em disconnected}
\cite{mo2:pre96}. When the value of $\ep$ is decreased, one should
find coexistence of both spot and stripe morphologies in the patterns
formed as a result of the destabilization of a spot
(Fig. \ref{f:split}). So, all the processes associated with the
dynamics of the interfaces: fission, elongation, tip splitting,
fingering, and wriggling \cite{seul95}, should generally be important
for the evolution of a single unstable spot.

\subsection{Coarsening and disorder} \label{s:coarse}

If the system is quenched deeply into the region in which the
homogeneous phase is unstable, at first a small-scale multidomain
pattern will form. Since the effect of the long-range interaction can
be seen only on the length scale $R \gtrsim \ep^{-2/3}$, initially the
long-range interaction will be negligible. Therefore, immediately
following the quench the pattern will undergo transient {\em
coarsening}. Note that in systems with the non-conserved order
parameter this transient coarsening may proceed at arbitrary $\phb$,
since the long-range interaction ensures the conservation of the total
amount of the order parameter. For example, in the case of
Eq.~(\ref{dyn}) the interfaces of the domains will be driven by
curvature subject to global coupling, so the characteristic radius of
the domains will obey the standard $t^{1/2}$ law independently of the
volume fraction \cite{m:phd,meerson96,rubinstein92}. As a result, the
characteristic size of the domains will grow until it becomes
comparable with the equilibrium size of Eq.~(\ref{R:gen}).  Then the
long-range interaction will stabilize the pattern, so at some point
the coarsening will become {\em arrested} (see also
\cite{bahiana90,glotzer95}).  This scenario is observed in the
experiments on thin diblock copolymer films \cite{coulon93}. Note that
this coarsening can be viewed as the consequence of the repumping
instability discussed in Secs. \ref{s:fluct} and \ref{s:perstab}.

When the temperature in the mean-field model from Sec. \ref{s:mf} is
slowly lowered, so that the value of $|\phb|$ becomes lower than
$|\phb_c|$, the homogeneous state becomes unstable with respect to the
fluctuations with the wave length $\lambda \sim \ep^{-1/2}$ (see
Sec. \ref{s:mf}). Thus, at the threshold of the instability the domain
pattern with characteristic size $\sim \lambda$ will start to
form \cite{ko:book,mo2:pre96}. These domains will still be smaller
than the equilibrium size $R \sim \ep^{-2/3}$, so the instability will
be followed by coarsening, just like in the case of the renormalized
model and in the mean-field model not close to $\phb_c$.

The results of the simulations of Eqs. (\ref{dyn}) displaying
transient coarsening are presented in Fig. \ref{f:coarse1}.  In all
these simulations the initial conditions were taken as $\phi = \phb$
plus small random noise. One can see that the morphology of the
pattern is determined by the volume fraction of the positive
domains. When $\phb = 0$ the pattern that forms in the end of the
simulation is a bicontinuous domain patterns similar to patterns
forming in the process of Ostwald ripening after the critical quench
\cite{bray94}.  When there is a small asymmetry between the positive
and negative domains [Fig. \ref{f:coarse1}(b)], the pattern in the end
of the simulation looks like a collection of disconnected spots and
stripes of different shapes and sizes. When the asymmetry between the
positive and negative domains is strong [Fig. \ref{f:coarse1}(c)],
only the spot morphology survives, and in the end the pattern is made
of a polydisperse mixture of spots.

Let us emphasize that the patterns that form at the end of the
simulations of Fig. \ref{f:coarse1} do not change in time, that is,
they are metastable. Each of these patterns is completely disordered,
and is {\em in no way} close to the perfectly ordered patterns that
are expected to be the global minimizers of the free energy. In the
absence of the noise the shape of the pattern at long times is
determined only by the random initial conditions.  The numerical
analysis of Eqs. (\ref{dyn}) shows that by changing the random seed
which determines the initial condition at the start of the simulation
one will get totally different metastable patterns in the end, so the
system is in fact very sensitive to the initial conditions. This also
suggests that besides the ordered equilibrium patterns, there exist a
huge number of irregular metastable patterns which locally minimize
the free energy. Thus, a typical pattern that should form as a result
of the fast quench must be disordered.

\begin{figure}
\centerline{\psfig{figure=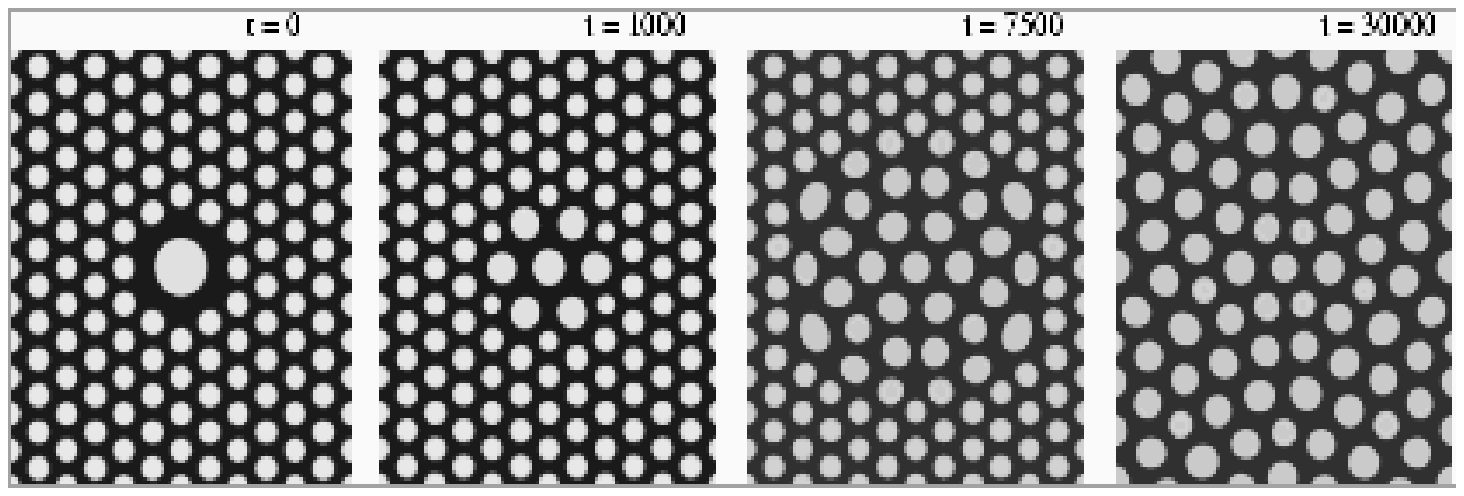,width=6in}}
\caption{The effect of a large-scale fluctuation on a hexagonal
pattern. Result of the numerical solution of Eq.~(\ref{dyn}) with $\ep
= 0.025$, $\phb = -0.2$, and periodic boundary conditions. The system
is $400 \times 460$.}
\label{f:relax}
\end{figure}

Once the metastable equilibrium is achieved, the patterns will evolve
by thermally activated processes. Indeed, in order for a pattern with
lower free energy to form, some of the domains may have to disappear
and some may have to be created, since the connectivity of different
metastable domain pattern is not generally the same. This requires to
overcome large free energy barriers $\Delta F \propto
\ep^{-2/3}$. Once again, on the time scale $\tau_\mathrm{nucl} \sim
e^{c \Delta F}$ the system will enter the aging regime. To mimic this
situation, we performed a numerical simulation of Eq.~(\ref{dyn}) with
a special initial condition in the form of a metastable hexagonal
pattern with a single bigger spot in the center
(Fig. \ref{f:relax}). As time goes on, the pattern tries to adjust to
accommodate a defect it is presented with. Let us emphasize that,
according to Fig. \ref{f:relax}, the defect {\em propagates} to
distances much larger than the characteristic size of the domains. In
the end, the pattern becomes completely disordered, with no traces of
the original hexagonal ordering.

Let us now ask a different question. Suppose that we already have an
equilibrium domain pattern in the system. What happens if at some
moment the temperature of the system is raised or lowered? This
question is related to what happens if the system is gradually cooled
below the transition temperature. Suppose the system is initially
occupied by the lowest energy hexagonal pattern (at a given
temperature). When the temperature of the system is lowered, the
values of $\ep$ and $|\phb|$ will decrease, so this pattern will no
longer correspond to the equilibrium pattern. To see this, let us
write down the length scales in Eq.~(\ref{hier:gen}) in the original
(unscaled) variables. Using the mean-field scaling \cite{landau5} and
the definition of $\ep$ from Eq.~(\ref{ep}), for example, we obtain
\begin{eqnarray}
l \sim |\tau|^{-1/2}, ~~~~~~~\lambda \sim \al^{-1/4}, \nonumber \\ R
\sim \al^{-1/3} |\tau|^{1/6}, ~~L \sim \al^{-1/2} |\tau|^{1/2}.
\end{eqnarray}
One can see from here that if the temperature is abruptly lowered, the
equilibrium size $R$ will increase. At the same time, the {\em
physical} size of the pattern will remain the same, so the relative
size of the pattern will decrease with respect to the new value of
$R$.

If the temperature drop is sufficiently small, the pattern will remain
metastable (see Sec. \ref{s:perstab}). However, when the temperature
falls below a certain critical temperature, the pattern will become
{\em unstable} with respect to repumping (Sec. \ref{s:perstab}). The
repumping will lead to the collapse of a fraction of the domains and
growth of the rest, so effectively, this will be equivalent to the
increase of the characteristic interdomain distance. The resulting
pattern will again be metastable. Note however, that it will
necessarily become disordered, since randomness is involved in the
destabilization of the hexagonal pattern. When the temperature gets
lower, the metastable pattern will again destabilize, and produce a
new metastable pattern with a greater characteristic domain size. This
process will go on. Thus, we will have a stepwise relaxation process
creating disordered patterns. Similar effect will be realized if one
takes a hexagonal pattern as an initial condition and gradually raises
the temperature. At some moment the pattern will become unstable with
respect to the asymmetric deformations, so domains of complex shapes
will start to form, thus effectively making the distance between the
patterns smaller. Such a metastable domain pattern will further
destabilize at higher temperature. These conclusions are confirmed by
the numerical simulations of Eqs. (\ref{dyn}) \cite{m1:prl97}. Note
that these arguments imply that disordered patterns will form even as
a result of a slow (but fast compared to $\tau_\mathrm{nucl}$) quench
below the transition temperature. All this indicates that disorder is
an intrinsic state of the systems with long-range competing
interactions. This is also seen in the experiments
\cite{cape71,coulon93,heckl86,seul95}.

Similarly, when one starts with a lamellar pattern and raises the
temperature, the pattern will become unstable with respect to
wriggling (see Sec. \ref{s:perstab}).  If the temperature is further
increased, the corrugation instability and fingering will
follow. Notice that in contrast to the hexagonal patterns, the
lamellar pattern will always remain metastable when the temperature is
lowered, since there is no repumping instability in this case. This
means that a metastable lamellar pattern is more likely to survive
after a slow (but fast compared with $\tau_\mathrm{nucl}$) critical
quench.

\section{Conclusions} 

In this work we have presented an energetic approach to the study of
inhomogeneous states (patterns) in systems with competing short-range
attractive and long-range repulsive Coulomb interactions.  Our
approach becomes universal for systems with weak Coulomb interaction
in the vicinity of the microphase separation transition, thus allowing
to treat a variety of physical situations which involve competing
Coulomb interactions.

By the very definition of the domain patterns, the width of the domain
wall should be much smaller than the characteristic size of an
individual domain. This requires that the Coulomb interaction be
sufficiently weak in order for these patterns to be feasible. On the
other hand, one can take advantage of this and study these patterns in
the asymptotic limit of infinitely weak Coulomb interaction. This
poses a challenge, however, since this interaction is a singular
perturbation to the short-range interaction.

We have performed an asymptotic analysis of the free energy in the
limit of vanishingly small strength of Coulomb interaction ($\ep
\rightarrow 0$). Our main finding is that in this limit the energetics
of the patterns are described by the locations of the domain
interfaces. In fact, an important hierarchy of length scales appears
in the system [Eq.~(\ref{hier:gen})]. Our second major observation is
that the characteristic size of the domains in a stable domain pattern
has to scale as $\ep^{-2/3}$. This is different from the similar
estimates based on the properties of {\em global} minimizers of the
free energy \cite{muller93,onishi99,ohta86,lorenzana01}. What we
showed in Sec. \ref{s:fluct} in general and Secs. \ref{s:morsol} and
\ref{s:perstab} for particular patterns is that unless this scaling is
obeyed, the pattern cannot be a {\em local} minimizer and thus
thermodynamically stable.

In our analysis, the starting point was the mean-field free energy
functional from Eq.~(\ref{FF}). We chose to perform our calculations
using Eq.~(\ref{F}) for two reasons. First, this is a universal
functional that is obtained in the vicinity of the microphase
separation transition and therefore may be applied to a variety of
systems. Second, using this functional we could obtain very explicit
results, making our presentation more tractable. It is not difficult
to see that all our calculations can be extended to the more general
functional from Eq.~(\ref{FF}). The only difference is that in the
case of Eq.~(\ref{FF}) the ``positive'' and ``negative'' domains may
have asymmetric linear response coefficients $\kappa_\pm$ instead of a
single $\kappa$ in the case of Eq.~(\ref{F}). Nevertheless, in the
case of Eq.~(\ref{FF}) we can choose $\kappa = \kappa_+ f + \kappa_-
(1 - f)$, where $f$ is the volume fraction of the ``positive'' domains
(similar ideas were used in   \cite{m1:pre97}). Indeed, since
$\kappa$ is responsible for screening, we can average the response of
the order parameter on the scale of the domains, which is much smaller
than the screening length (Sec. \ref{s:mf}). The new definition of
$\kappa$ also takes into account that to the leading order the
(locally) averaged value of the volume fraction $f$ is independent of
space. The latter can be easily seen from the analog of
Eq.~(\ref{inh}) obtained from Eq.~(\ref{FF}), if one integrates this
equation over a closed volume of size $\ep^{-2/3} \ll \ell \ll
\ep^{-1}$, uses the Gauss theorem, and takes into account that $\phi$
is nearly constant in the domains and $|\nabla \psi| \sim \psi / R
\sim \ep^{4/3}$ (see Sec. \ref{s:int}).

Let us point out that the interfacial representation of the free
energy given by Eq.~(\ref{F:inter}) which we obtained from the free
energy functional in Eq.~(\ref{F}) in the asymptotic limit $\ep
\rightarrow 0$ may in fact itself form a basis for studying the domain
patterns in systems with long-range Coulomb interactions, provided
that the driving force for the formation of these patterns is the
competition of the Coulomb energy with the surface energy (see, for
example, \cite{nagaev95,lorenzana01,bates99}). In this formulation our
results can be applied to an even wider range of systems, which may
not generally possess a free energy functional, like the one in
Eq.~(\ref{F}). For example, our asymptotic results should apply to the
ferromagnetic nearest-neighbor Ising models frustrated by Coulomb
interactions \cite{emery93,grousson01,viot98,low94,schmalian00}. We
argued that in these systems thermal fluctuations should only
renormalize the effective coupling constant of the Coulomb
interactions without qualitatively affecting the overall picture
(Sec. \ref{s:renorm}). These predictions are difficult to compare with
the recent Monte-Carlo simulations \cite{viot98,grousson01} because of
the limitation of the latter on the system's size. Nevertheless, the
result of \cite{viot98} about the location of the microphase
separation transition, which gives $\tau_c \sim \al^x$, with $x \simeq
0.25-0.35$, is not far from our prediction from Sec. \ref{s:renorm} of
$\tau_c \sim \al^{0.40}$ for the three-dimensional Ising model. Note
that we do not expect to find an avoided critical behavior discussed
in the context of the mean-spherical models \cite{chayes96}.

An interesting question arising in systems with long-range competing
interactions is the nature of the thermodynamic ground state below the
microphase separation transition temperature. We emphasize that our
stability analysis of stationary patterns only addresses small-scale
thermal fluctuations, so we are really talking about {\em
metastability} of these patterns. At the same time, rare large-scale
thermal fluctuations may lead to nucleation or transitions between
different metastable patterns (see Secs. \ref{s:nucl} and
\ref{s:coarse}). In this sense, if there are enough metastable
patterns, the global minimizer of the interfacial free energy, which
is presumably a highly symmetric periodic pattern \cite{ohta86}, has
little to do with the thermodynamic ground state of the system.

In fact, we see that the stationary metastable patterns that form in
one way or another are typically highly disordered
(Sec. \ref{s:coarse}). Although the basic interaction between
different domains is repulsion, the domains rarely arrange themselves
in a close-packed fashion. The reason for that is that even though the
interaction between the domains is repulsive, the range of this
interaction, which is determined by the screening length $L \sim
\ep^{-1}$ is much greater than the characteristic interdomain distance
$R \sim \ep^{-2/3}$. So, a single domain interacts simultaneously with
many other domains and not only with its nearest neighbors. Therefore,
the optimization of the free energy becomes a collective problem, and
a huge number of disordered metastable states appears. Then, the
configurational entropy of these metastable disordered states may
overwhelm their energy disadvantage \cite{bouchaud97,chowdhury}.

Furthermore, a large-scale fluctuation whose size is comparable with
the domain size $\sim R$ may propagate its action to the much larger
distance $\sim L$ (see Fig. \ref{f:relax}). It would seem natural to
expect that even if the system is in the lowest energy state, a
sufficiently strong fluctuation will frustrate a region much larger
than the size of such a fluctuation, which may lead to increase of the
degree of disorder with time. In this sense systems with long-range
competing Coulomb interactions can be considered as {\em structural
glasses} \cite{m:phd,emery93}.  We emphasize that in these systems the
disorder is self-induced. As we showed in Sec. \ref{s:coarse}, these
systems can age on very long time scales and exhibit complex
relaxation phenomena even in the case when the equations of motion for
the patterns are very simple. Note that in a recent paper, Schmalian
and Wolynes came to similar conclusions on the basis of their replica
analysis of Eq.~(\ref{F}) treated as an effective hamiltonian
\cite{schmalian00}. Their calculations suggest that the number of
metastable states grows exponentially with the system's volume,
leading to an {\em ideal glass transition} below the microphase
separation transition temperature. Also note that spin systems
frustrated by Coulomb interactions were proposed for studying glassy
behavior in the supercooled liquids \cite{kivelson95}.

\section{acknowledgments} 

The author would like to acknowledge many valuable discussions with
E. Demler, W. Klein and V. V. Osipov. The algebraic calculations of
the paper benefited from the use of Mathematica 4.0 software.

\appendix

\section{Free energy} \label{a:free}

Here we present the details of our manipulations of the interfacial
free energy from Eq.~(\ref{F:inter}) and the effective field $\psi$
from Eq.~(\ref{sm}). Let us first show the derivation of
Eq.~(\ref{F:int}) from Eq.~(\ref{longcontr}). In view of
Eq.~(\ref{sm}), we have
\begin{eqnarray} \label{A1}
F_\mathrm{long-range} = \frac{1}{2} \int d^d x (\phi_\mathrm{sh} -
\phb) \psi.
\end{eqnarray}
According to Eq.~(\ref{sm})
with $\phi_\mathrm{sh} = \pm 1$, we have
\begin{eqnarray} \label{A2}
\psi = - \ep^2 (1 + \phb) \int d^d x' G_\ep(x - x') \nonumber \\ + 2
\ep^2 \int_{\Omega_+} d^d x' G_\ep(x - x'),
\end{eqnarray}
where the first integral is over the whole space. This integral is
equal to $1/(\ep \kappa)^2$, according to
Eq.~(\ref{Gsc}). Substituting this back into Eq.~(\ref{A1}), after
simple algebra we arrive at Eq.~(\ref{F:int}).

Let us now derive Eq.~(\ref{F:inter}) from Eq.~(\ref{F:int}). Using
Eq.~(\ref{Gsc}) and applying Gauss's theorem, we calculate the
long-range term:
\begin{eqnarray}
2 \ep^2 \int_{\Omega_+} \int_{\Omega_+} d^d x \, d^d x' G_\ep(x - x')
= \nonumber \\ {2 \over \kappa^2} \int_{\Omega_+} \int_{\Omega_+} d^d
x d^d x' \left( \delta^{(d)}(x - x') + \nabla^2 G_\ep(x - x') \right)
= \nonumber \\ {2 \over \kappa^2} \int_{\Omega_+} d^d x + {2 \over
\kappa^2} \int_{\Omega_+} d^d x' \oint dS \{\hat n \cdot \nabla
G_\ep(x - x') \} = \nonumber \\ {2 \over \kappa^2 d} \oint dS (\hat n
\cdot \vec x) \nonumber \\ - {2 \over \kappa^2} \oint dS
\int_{\Omega_+} d^d x' \nabla' \cdot \{ \hat n G_\ep(x - x') \} =
\nonumber \\ {2 \over \kappa^2 d} \oint dS (\hat n \cdot \vec x) - {2
\over \kappa^2} \oint dS \oint dS' (\hat n \cdot \hat n') G_\ep(x -
x'). \nonumber \\
\end{eqnarray}

Let us now derive Eq.~(\ref{psicont}). Using Eq.~(\ref{Gsc}),
Eq.~(\ref{A2}), and Eq.~(\ref{G}) to express the $\delta$-function in
terms of $G$, we get
\begin{eqnarray}
\psi = -{1 + \phb \over \kappa^2} \nonumber \\ + {2 \over \kappa^2}
\int_{\Omega_+} d^d \, x' \nabla^2 \{ G_\ep(x - x') - G(x - x') \} =
\nonumber \\ -{1 + \phb \over \kappa^2} + {2 \over \kappa^2} \oint dS'
\{ \hat n' \cdot \nabla' (G_\ep - G) \},
\end{eqnarray}
where we applied the Gauss's theorem.

Now let us calculate the first and second variations of the
interfacial free energy. Let $\rho(x)$ be a normal displacement of the
interface at point $x$ on the interface, which is positive if the
displacement is into the positive domain and vice versa. Note that
according to our definition, $\rho > 0$ corresponds to shrinking of
$\Omega_+$.

Up to second order in $\rho$, the change of the surface free energy
$\Delta F_\mathrm{surf}$ is given by a well-known formula (see, for
example,   \cite{taylor3}):
\begin{eqnarray} \label{A4}
\Delta F_\mathrm{surf} = - 2 \sigma_0 \oint dS \, H \rho + \nonumber
\\ {\sigma_0 \over 2} \oint dS \left( |\nabla_\perp \rho|^2 + 2 K \rho^2
\right),
\end{eqnarray}
where $H$ and $K$ are mean and Gaussian curvatures at a given point of
the interface, respectively, and $\nabla_\perp$ denotes the gradient
restricted to the interface. The mean curvature is positive if the
positive domain is convex. The change of the long-range contributions
to the free energy is given by an integral over a thin layer of
thickness $\rho$ over the interface. According to Eq.~(\ref{F:int}),
we have
\begin{eqnarray}
\Delta F_\mathrm{long-range} = \nonumber \\ {2 (1 + \phb) \over
\kappa^2} \oint dS \int_0^{\rho(x)} dz (1 - 2 H z + K z^2) \nonumber
\\ - 4 \ep^2 \oint dS \int_0^{\rho(x)} dz (1 - 2 H z + K z^2) \times
\nonumber \\ \int_{\Omega_+} d^dx' \, G_\ep(x - \hat n z - x')
\nonumber \\ + 2 \ep^2 \oint dS \int_0^{\rho(x)} dz (1 - 2 H(x) z +
K(x) z^2) \times \nonumber \\ \oint dS' \int_0^{\rho(x')} dz' (1 - 2
H(x') z' + K(x') {z'}^2) \times \nonumber \\ G_\ep(x - \hat n z - x' +
\hat n' z'),
\end{eqnarray}
where we used the fact that with our definition of the sign of
principal curvatures $d^d x = (1 - k_1 z) (1 - k_2 z) dz dS = (1 - 2 H
z + K z^2) dz dS$ at a point distance $z$ away from the
interface. Retaining only the terms up to second order in $\rho$ and
using Eq.~(\ref{A2}), we obtain
\begin{eqnarray} \label{A6}
\Delta F_\mathrm{long-range} = - 2 \oint dS \, \psi \rho \nonumber \\
+ 2 \oint dS \, H \psi \rho^2 + \oint dS \, (\hat n \cdot \nabla \psi)
\rho^2 \nonumber \\ + 2 \ep^2 \oint dS \oint dS' G_\ep(x - x') \rho(x)
\rho(x'),
\end{eqnarray}
where we expanded $\psi$ in the Taylor series in $z$. From this, and
Eq.~(\ref{A4}), we get
\begin{eqnarray}
\delta F = -2 \oint dS ( \sigma_0 H + \psi) \rho,
\end{eqnarray}
so the critical points must satisfy Eq.~(\ref{v}). Similarly, using
Eq.~(\ref{v}) in Eq.~(\ref{A6}), we obtain
\begin{eqnarray}
\delta^2 F = \sigma_0 \oint dS \left( |\nabla_\perp \rho|^2 + 2 K
\rho^2 \right) \nonumber \\ + \oint dS \{ 2 ( \hat n \cdot \nabla
\psi) - 4 \sigma_0 H^2 \} \rho^2 \nonumber \\ + 4 \ep^2 \oint dS \oint
dS' G_\ep(x - x') \rho(x) \rho(x'),
\end{eqnarray}
which in view of Eq.~(\ref{psicont}) coincides with Eq.~(\ref{matr}). 

\section{Optimal period of hexagonal and lamellar patterns}
\label{a:hexlam} 

Here, we give the Wigner-Seitz calculation of the period of the
hexagonal and lamellar patterns. 

\subsection{Hexagonal pattern}

We start with a hexagonal pattern. Consider Eq.~(\ref{psiscreened}) on
a disk of radius $\rs = 3^{1/4} \ls_p / \sqrt{2 \pi}$, with no flux
boundary conditions. Neglecting the term $\ep^2 \kappa^2 \psi$ and
using Eq.~(\ref{fraction}), we write
\begin{widetext}
\begin{eqnarray}
{d^2 \psi \over d r^2} + {1 \over r} {d \psi \over dr} + \ep^2 \{
\theta(\rs_s - r) - \theta(r - \rs_s) - 2f + 1 \} = 0,
\end{eqnarray}
where $r$ is the radial coordinate and $\theta(x)$ is the Heaviside
step. The solution of this equation that satisfies Eq.~(\ref{v}) is
given by
\begin{eqnarray} \label{B2}
\psi = \left\{
\begin{array}{lc}
\frac{1}{2} \ep^2 \{ (f - 1) r^2 + \rs^2 f (1 - f) \} - \frac{1}{2}
\sigma_0 \rs^{-1} f^{-1/2}, & 0 \leq r \leq \rs_s, \\ \\ \frac{1}{2}
\ep^2 \{ f r^2 - f^2 \rs^2 + f \rs^2 (\ln f \rs^2 - 2 \ln r) \} -
\frac{1}{2} \sigma_0 \rs^{-1} f^{-1/2}, & \rs_s \leq r \leq \rs.
\end{array}
\right.
\end{eqnarray}
\end{widetext}
where we took into account that $f = \rs_s^2 / \rs^2$. According to
Eq.~(\ref{longcontr}) with $\phi_\mathrm{sh} = \pm 1$ and
Eq.~(\ref{A2}), the long-range contribution to the free energy can be
computed as
\begin{eqnarray}
F_\mathrm{long-range} = \nonumber \\ (1 - f) \int_{\Omega_+} \psi d^d
x - f \int_{\Omega_-} \psi d^d x = \nonumber \\ 2 \pi (1 - f)
\int_0^{\rs_s} r \psi(r) dr - 2 \pi f \int_{\rs_s}^\rs r \psi(r) dr.
\end{eqnarray}
Combining this with the surface energy $F_\mathrm{surf} = 2 \pi
\sigma_0 \rs_s$ and using Eq.~(\ref{B2}), we get that the free energy
per unit area is
\begin{eqnarray}
{F \over \pi \rs^2} = {2 \sigma_0 f^{1/2} \over \rs} + {\ep^2 \rs^2
f^2 \over 2} (f - 1 - \ln f),
\end{eqnarray}
so, minimizing this expression with respect to $\rs$ with fixed $f$,
we obtain that the minimum is attained at $\rs = \rs^*$, where
\begin{eqnarray}
\rs^* = \ep^{-2/3} f^{-1/2} \left( {2 \sigma_0 \over f - \ln f - 1}
\right)^{1/3}.
\end{eqnarray}
Using the definition of $\rs$ in terms of $\ls_p$, this equation is
rewritten as Eq.~(\ref{lp2eq}).

\subsection{Lamellar pattern}

Similarly, for the lamellar pattern centered at $x = 0$ we get
\begin{widetext}
\begin{eqnarray}
\psi= \left\{
\begin{array}{lc}
\ep^2 (f - 1) x^2 + \frac{1}{4} \ep^2 f^2 (1 - f) \ls_p^2, & 0 \leq x
\leq \ls_s/2, \\ \\ \ep^2 f x^2 - \ep^2 f \ls_p x + \frac{1}{4} \ep^2
f^2 ( 2 - f) \ls_p^2, & \ls_s/2 \leq x \leq \ls_p/2,
\end{array}
\right.
\end{eqnarray}
\end{widetext}
where we used the fact that $f = \ls_s / \ls_p$. Calculating the free
energy per unit length, we get
\begin{eqnarray}
{F \over \ls_p} = {2 \sigma_0 \over \ls_p} + {1 \over 6} \ep^2 \ls_p^2
f^2 (1 - f)^2.
\end{eqnarray}
Minimizing this expression with respect to $\ls_p$, we obtain
Eq.~(\ref{lp1eq}). 

\section{Stability of hexagonal and lamellar patterns}

Here we present the details of our calculations of
Eqs.~(\ref{disp:hex}) and (\ref{disp:lam}). 

\subsection{Hexagonal pattern} \label{a:hexstab}

We begin with the hexagonal pattern. We define
\begin{eqnarray}
\mathsf R_{mm'}(\mathbf k) = {2 \ep^2 \rs_s \over \pi} \sum_n
\int_0^{2 \pi} d \varphi \int_0^{2 \pi} d \varphi' && \nonumber \\ &&
\hspace{-6cm} \times e^{i m \varphi - i m' \varphi' + i \mathbf k
\cdot \mathbf R_n} G_\ep(\mathbf R_n + \mathbf r(\varphi') - \mathbf
r(\varphi)),
\end{eqnarray}
where $\mathbf r(\varphi) = (\rs_s \cos \varphi, \rs_s \sin \varphi)$
and the summation is over the lattice: $\mathbf R_n = n_1 \mathbf a_1
+ n_2 \mathbf a_2$, where $\mathbf a_1 = \frac{1}{2} \ls_p(\sqrt{3},
1)$ and $\mathbf a_2 = \frac{1}{2} \ls_p (\sqrt{3}, -1)$. Using the
Fourier representation of $G_\ep(x - x')$, we obtain
\begin{eqnarray}
\mathsf R_{mm'}(\mathbf k) = {2 \ep^2 \rs_s \over \pi} \sum_n
\int_0^{2 \pi} d \varphi \int_0^{2 \pi} d \varphi' e^{i m \varphi - i
m' \varphi' } && \nonumber \\ &&
\hspace{-7cm} \times \int {d \mathbf k' \over (2 \pi)^2} {e^{i \mathbf
k \cdot \mathbf R_n - i \mathbf k' \cdot (\mathbf R_n + \mathbf
r(\varphi') - \mathbf r(\varphi)) } \over |\mathbf k'|^2 + \ep^2
\kappa^2}
\label{C1} \\ &&
\hspace{-7cm} = {2 \ep^2 \rs_s \over \pi v} \sum_n \int_0^{2 \pi} d
\varphi \int_0^{2 \pi} d \varphi' e^{i m \varphi - i m' \varphi'}
\nonumber \\ && \hspace{-7cm} \times {e^{i (\mathbf k + \mathbf k_n)
\cdot (\mathbf r(\varphi) - \mathbf r(\varphi'))} \over |\mathbf k +
\mathbf k_n|^2 + \ep^2 \kappa^2},
\end{eqnarray}
where $v = \ls_p^2 \sqrt{3}/2$ is the area of the Wigner-Seitz cell,
the sum is now over the reciprocal lattice, and we took into account
that $\sum_n e^{i (\mathbf k - \mathbf k') \cdot \mathbf R_n} = (4
\pi^2 /v) \sum_n \delta(\mathbf k' - \mathbf k - \mathbf k_n)$. 

We proceed:
\begin{eqnarray}
\mathsf R_{mm'}(\mathbf k) = {2 \ep^2 \rs_s \over \pi v} \sum_n {1
\over |\mathbf k + \mathbf k_n|^2 + \ep^2 \kappa^2} \times \nonumber
\\ \int_0^{2 \pi} d \varphi e^{i m \varphi + i (\mathbf k + \mathbf
k_n) \cdot \mathbf r(\varphi)} \int_0^{2 \pi} d \varphi' e^{- i m'
\varphi' - i (\mathbf k + \mathbf k_n) \cdot \mathbf r(\varphi')}
\nonumber \\ = {2 \ep^2 \rs_s \over \pi v} \sum_n {1 \over |\mathbf k
+ \mathbf k_n|^2 + \ep^2 \kappa^2} \times \nonumber \\ \int_0^{2 \pi}
d \varphi e^{i m \varphi + i |\mathbf k + \mathbf k_n | \rs_s
\cos(\varphi - \vartheta_{\mathbf k + \mathbf k_n})} \times \nonumber
\\ \int_0^{2 \pi} d \varphi' e^{- i m' \varphi' - i |\mathbf k +
\mathbf k_n| \rs_s \cos (\varphi' - \vartheta_{\mathbf k + \mathbf
k_n})} \nonumber \\ = {8 \pi \ep^2 \rs_s \over v} \sum_n {e^{i(m - m')
\vartheta_{\mathbf k + \mathbf k_n}} \over |\mathbf k + \mathbf k_n|^2
+ \ep^2 \kappa^2} \times \nonumber \\ i^m J_m(|\mathbf k + \mathbf
k_n| \rs_s) i^{-m'} J_{m'} (|\mathbf k + \mathbf k_n| \rs_s), \nonumber
\\
\end{eqnarray}
where we introduced the angle $\vartheta_{\mathbf k + \mathbf k_n}$
between the vector $\mathbf k + \mathbf k_n$ and the $x$-axis and used
the integral representation of the Bessel function. After a few
algebraic manipulations, this equation can be converted to
Eq.~(\ref{Rmmk}). Using the reciprocal lattice vectors $\mathbf b_1 =
2 \pi \ls_p^{-1} (1/\sqrt{3}, 1)$ and $\mathbf b_2 = 2 \pi \ls_p^{-1}
(1/\sqrt{3}, -1)$, so $\mathbf k_n = n_1 \mathbf b_1 + n_2 \mathbf
b_2$, this sums can be evaluated numerically by truncating the
summation at sufficiently large $|n_1|$ and $|n_2|$.

An alternative representation for $\mathsf R_{mm'}(\mathbf k)$, which
allows to explicitly calculate its diagonal elements, can be obtained
by performing the summation in real space, rather than over the
reciprocal lattice. We rewrite Eq.~(\ref{C1}) as
\begin{eqnarray}
\mathsf R_{mm'}(\mathbf k) = 4 \ep^2 \rs_s \sum_n e^{i \mathbf k \cdot
\mathbf R_n} \int_0^\infty {q d q \over q^2 + \ep^2 \kappa^2} \times
\nonumber \\ \int_0^{2 \pi} {d \vartheta \over 2 \pi} e^{-i q |\mathbf
R_n| \cos(\vartheta - \vartheta_n)} \times \nonumber \\ \int_0^{2 \pi}
{d \varphi \over 2 \pi} e^{i m \varphi + i q \rs_s \cos(\varphi -
\vartheta)} \times \nonumber \\ \int_0^{2 \pi} {d \varphi' \over 2
\pi} e^{-i m' \varphi' - i q \rs_s \cos(\varphi' - \vartheta)}
\nonumber \\ = 4 \ep^2 \rs_s \sum_n e^{i \mathbf k \cdot \mathbf R_n}
\int_0^\infty { q dq \over q^2 + \ep^2 \kappa^2} \times \nonumber \\
\int_0^{2 \pi} {d \vartheta \over 2 \pi} e^{-i q | \mathbf R_n|
\cos(\vartheta - \vartheta_n) + i (m - m') \vartheta} \times \nonumber
\\ i^{m-m'} J_m(q \rs_s) J_{m'}(q \rs_s), ~~~
\end{eqnarray}
where $\vartheta_n$ is the angle between $\mathbf R_n$ and the
$x$-axis. Calculating the integral over $\vartheta$, we obtain
\begin{eqnarray}
\mathsf R_{mm'}(\mathbf k) = 4 \ep^2 \rs_s \sum_n e^{i \mathbf k \cdot
\mathbf R_n + i(m - m') \vartheta_n} \times \nonumber \\ \int_0^\infty
{q d q \over q^2 + \ep^2 \kappa^2} J_{m-m'}(q |\mathbf R_n|) J_m(q
\rs_s) J_{m'}(q \rs_s).  \label{C6}
\end{eqnarray}

In calculating $\mathsf R_{mm'}(\mathbf k)$, to the leading order in
$\ep$ one can neglect the term $\ep^2 \kappa^2$ in the denominator of
Eq.~(\ref{Rmmk}) or Eq.~(\ref{C6}). Setting $\ep$ to zero, we can
calculate the diagonal elements $\mathsf R_{mm} (\mathbf k)$ for $m
\geq 2$. After some algebra
\begin{eqnarray} \label{C7}
\mathsf R_{mm}(\mathbf k) = {2 \ep^2 \rs_s \over m},
\end{eqnarray}
where we took into account that the integrals in Eq.~(\ref{C6}) all
vanish for $\mathbf R_n \not = \mathbf 0$. Thus, the diagonal elements
$\mathsf R_{mm}(\mathbf k)$ are independent of $\mathbf k$ and coincide
with those of a single spot.

Caution, however, is necessary when $|\mathbf k| \lesssim \ep$. In
this case the $\mathbf k_n = \mathbf 0$ contribution to the sum in
Eq.~(\ref{Rmmk}) will be singular for $m, m' = 0, \pm 1$. Taking only
the contribution of $\mathbf k_n = \mathbf 0$, for $|\mathbf k| \ll 1$
we obtain
\begin{eqnarray} \label{C8}
\mathsf R_{0,0} (\mathbf k) = {16 \pi \ep^2 \rs_s \over \ls_p^2
\sqrt{3}} \left( {1 \over |\mathbf k|^2 + \ep^2 \kappa^2} \right),
\end{eqnarray}
where we expanded the Bessel functions in the Taylor series and
retained only the leading term. Now, to calculate $\mathsf
R_{1,1}(\mathbf k)$ for $\mathbf k = \mathbf 0$, note that if one
formally sets $\ep = 0$ in Eq.~(\ref{Rmmk}) with $m = m' = 1$, one
should get the result of Eq.~(\ref{C7}). On the other hand, for
$\mathbf k = \mathbf 0$ the term with $\mathbf k_n = \mathbf 0$ does
not contribute, while for other $\mathbf k_n$ the term $\ep^2
\kappa^2$ in Eq.~(\ref{Rmmk}) is a regular perturbation and can be
neglected. So to calculate $\mathsf R_{1,1}(\mathbf k)$ in the limit
$|\mathbf k| \rightarrow 0$, one has to subtract the $\ep \rightarrow
0$ limit of the $\mathbf k_n = \mathbf 0$ term in Eq.~(\ref{Rmmk})
from Eq.~(\ref{C7}). As a result, we obtain
\begin{eqnarray} \label{C9}
\mathsf R_{1,1}(\mathbf 0) = 2 \ep^2 \rs_s - {4 \pi \ep^2 \rs_s^3
\over \ls_p^2 \sqrt{3}},
\end{eqnarray}
where we expanded the Bessel functions in the Taylor series and
retained only the leading term. 

Since $\mathsf R_{0,0}(\mathbf k) \gg |\mathsf R_{mm'}(\mathbf k)|$
for $m, m' \not = 0$ and small $|\mathbf k|$, the $m = 0$ mode is the
eigenfunction of the operator $L$ in Eq.~(\ref{lin}) for vanishing
$|\mathbf k|$. The analysis of Eq.~(\ref{disp:hex}) with $m = m' = 0$
and $\mathsf R_{0,0}(\mathbf k)$ from Eq.~(\ref{C8}) shows that the
hexagonal pattern is stable with respect to the long-wave modulation
of the spots' radii, as long as $\ls_p$ is large enough. 

\subsection{Lamellar pattern} \label{a:lamstab}

Let us now turn to the lamellar pattern. Calculate the matrix elements
of $4 \ep^2 G_\ep(x - x')$ between the right (``+'') and left (``--'')
walls of the stripe in the zeroth period for a given modulation:
\begin{eqnarray}
\langle + | 4\ep^2 G_\ep | + \rangle = \langle - | 4\ep^2 G_\ep | -
\rangle \nonumber \\ = {2 \ep^2 \over k} \sum_{n=-\infty}^{+\infty}
e^{-k |\ls_p n| + i k_\parallel n \ls_p} \nonumber \\ = {2 \ep^2 e^{i
k_\parallel \ls_p} \left( e^{2 k \ls_p} - 1 \right) \over k \left(
e^{(k + i k_\parallel) \ls_p} - 1 \right) \left( e^{k \ls_p} - e^{i
k_\parallel \ls_p} \right)},
\end{eqnarray}
and
\begin{eqnarray}
\langle + | 4\ep^2 G_\ep | - \rangle = {\langle - | 4\ep^2 G_\ep | +
\rangle}^* \nonumber \\ = {2 \ep^2 \over k} \sum_{n=-\infty}^{+\infty}
e^{-k |\ls_p n - \ls_s | + i k_\parallel n \ls_p} \nonumber \\ = {2
\ep^2 e^{i k_\parallel \ls_p - k \ls_s} \over k} \left({e^{2 k \ls_s}
\over e^{k \ls_p} - e^{i k_\parallel \ls_p} } + {e^{k \ls_p} \over
e^{(k + i k_\parallel) \ls_p} - 1} \right), \nonumber \\
\end{eqnarray}
where we introduced $k = \sqrt{\ep^2 \kappa^2 + k_\perp^2}$, used the
fact that the Fourier transform of the Green's function $G_\ep$ in the
transverse direction is given by $\exp(-k |z - z'|)/2 k$, and summed
geometric series. After some algebra, the $2 \times 2$ matrix $\mathsf
R(k_\parallel, k_\perp)$ formed by these matrix elements can be
transformed into the following form
\begin{widetext}
\begin{eqnarray}
\mathsf R(k_\parallel, k_\perp)= {4 \ep^2 e^{k \ls_p} \over k \left( 1
- 2 e^{k \ls_p} \cos k_\parallel \ls_p + e^{2 k \ls_p} \right) }
\times \nonumber \\ \left(
\begin{array}{cc}
\sinh k \ls_p & \sinh [k(\ls_p - \ls_s)] + e^{i k_\parallel \ls_p}
\sinh k \ls_s \\ \sinh [k(\ls_p - \ls_s)] + e^{-i k_\parallel \ls_p}
\sinh k \ls_s & \sinh k \ls_p
\end{array}
\right).
\end{eqnarray}
\end{widetext}
This matrix can be easily diagonalized, after a few manipulations we
arrive at Eq.~(\ref{Rpmlam}). Then, Eq.~(\ref{l0:lam}) is obtained by
setting $\lambda_0 = -\mathsf R_-(0, 0)$ and taking only the leading
order terms. Note that for $k_\parallel = 0$ or $k_\parallel = \pi /
\ls_p$ the fluctuations corresponding to $\lambda_\pm$ are the
symmetric and antisymmetric deformations of stripes.

To obtain the energy of the long-wave distortions of the lamellar
pattern, we expand Eq.~(\ref{disp:lam}) with $\mathsf R_-$ into a
series in $k_\parallel$ and $k_\perp$ and retain the terms up to
quadratic in $k_\parallel$ and forth order in $k_\perp$. Then, to the
leading order in $\ep$, we get
\begin{eqnarray}
\lambda_- \simeq \frac{1}{2} f^2 (1 - f)^2 \ep^2 \ls_p^3 \,
k_\parallel^2 \nonumber \\ + \left( \sigma_0 - \frac{1}{6} f^2 (1 -
f)^2 \ep^2 \ls_p^3 \right) k_\perp^2 \nonumber \\ + \frac{1}{360}
\left[ f^2 (1 - f)^2 (1 + 2 f - 2 f^2) \ep^2 \ls_p^5 \right]
k_\perp^4,
\end{eqnarray}
where we used $\ls_s / \ls_p = f$. One can see that at $\ls_p =
\ls_p^*$ given by Eq.~(\ref{lp1eq}) the coefficient of $k_\perp^2$
changes sign from positive at $\ls_p < \ls_p^*$ to negative at $\ls_p
> \ls_p^*$, signifying an instability. At the same time, the
coefficient of $k_\perp^4$ is positive for all $0 < f < 1$.

Let us now discuss the stability of the lamellar patterns in one
dimension, which can be studied by looking at Eq.~(\ref{disp:lam})
with $k_\perp = 0$. Setting $k_\perp = 0$ and expanding in $\ep$,
after some algebra we obtain that to the leading order
\begin{eqnarray}
\lambda_- = {2 \ep^2 \ls_p \over 1 - \cos k_\parallel \ls_p} \biggl[ 1
- f + f^2 + f(1 - f) \cos k_\parallel \ls_p \nonumber \\ - \sqrt{1 - 2
f (1 - f)(1 - \cos k_\parallel \ls_p)} \bigg]. \nonumber \\
\end{eqnarray}
It is not difficult to verify that according to this equation
$\lambda_- \geq 0$ for all values of $k_\parallel$, so the lamellar
pattern is always stable regardless of the modulation vector
$k_\parallel$ in one dimension. This conclusion is applicable when
$\ln \ep^{-1} \ll \ls_p \ll \ep^{-1}$, when the assumptions of the
above equations are valid. Note that for $\ls_p$ outside this range
the one dimensional lamellar patterns (strata) may undergo a number of
instabilities \cite{ko:book}.

\bibliography{../main}

\begin{widetext}

\end{widetext}

\end{document}